\documentclass{jfm}

\usepackage{graphicx}
\usepackage{newtxtext}
\usepackage{newtxmath}
\usepackage{natbib}
\usepackage{hyperref}
\hypersetup{
    colorlinks = true,
    urlcolor   = blue,
    citecolor  = black,
}

\newcommand{\RomanNumeralCaps}[1]
\linenumbers

\title{Blast wave induced unsteady flow at the shock tube opening}

\author{Saini Jatin Rao\aff{1}, Akhil Aravind\aff{1}
 \and Saptarshi Basu\aff{1}
  \corresp{\email{sbasu@iisc.ac.in}}}
\affiliation{\aff{1}Department of Mechanical Engineering, Indian Institute of Science, Bangalore-560012, India}

\begin{document}
\maketitle

\begin{abstract}
Shock tubes have been a crucial device, facilitating studies across a wide range of practical applications.  An open-ended shock tube employing the wire-explosion technique with a rectangular cross section is used in the present study to generate blast waves over a Mach number range of $1.2–1.8$, enabling detailed investigation of unsteady compressible flow at the tube opening. The blast wave produces a complex flow field comprising a compressible vortex ring with a trailing jet, and several transient structures, including embedded shocks, inward-moving shock or reverse shocks, shear layers, and Prandtl–Meyer expansion fans. An approximate model based on a power-law density profile describes the blast evolution inside and outside the tube, with the equivalent source deduced from measured shock trajectories. The blast wave-tube exit interaction is analyzed using the method of characteristics with alternate exit boundary conditions. A steady-pressure outlet best reproduces experimental observations, predicting supersonic efflux, embedded shocks, expansion waves, and circulation production. Several previously unreported unsteady features, including reverse shock or "reshock" formation and embedded shock shedding, are documented. The findings highlight the intricate dynamics of various features associated with such highly transient, blast-driven flows emanating from an open-ended shock tube.

\end{abstract}



\section{Introduction}
\label{sec:Introduction}

Shock tubes have been instrumental in providing a canonical flow representation and a critical device, enabling studies spanning many practical applications \citep{kahaliEvolutionSecondaryFlow2020}.  The conventional shock tubes generally involve an impulsive "blow-down" of compressed air through a conduit, popularly known as a wind tunnel, generating high-speed flows and accompanying the namesake artifacts, i.e., the shock waves. This is typically achieved through a pressurized reservoir in the upstream and/or a depressurized (vacuum) reservoir in the downstream of the test section \citep{pope1965high}. This pressure contrast is established and held through a barrier, i.e., a diaphragm, which is ruptured manually to initiate the flow impulsively, forming a shock wave. This phenomenon generates other compressible features depending on a particular configuration, such as expansion fans, contact discontinuities, vortical structures, and other secondary shocks. The diagram-based shock tubes have been essential in the community, but suffer from a few drawbacks: (i) a huge footprint - they require ample space, large reservoirs, and long preparation times for a single trial, (ii) limited possibilities of transient flow profiles - in idealized conditions, they generate a steady flow until the reflected characteristics from the upstream boundary catches up with the leading shockwave. In a lot of practical applications involving shock-structure interaction, shock-induced atomization, or flame disruption, we rarely see a steady or uniform idealized flow. This motivates us to introduce the blast wave, another canonical discontinuous wave form of practical interest, typically originating from explosions \citep{korobeinikovGasDynamicsExplosions1971}.

Blast waves have been known for generations, but have been notoriously well-studied in secrecy during the Second World War, complementing the development of the first atomic bomb. This involved studying the implications of the blast wave produced by the explosion on the nearby infrastructure. Blast wave represents an unsteady shock wave, decaying in space and time. Any explosion with sufficiently large energy, impulsively deposited in a small region over a very short period, is accompanied by a blast wave propagating radially outward from the source. The attenuation is inherited by the unsteady wave front, with a decreasing speed as it moves away from the source. Unlike conventional unsteady shock motion, where the shock wave moves with a constant speed and the flow behind is uniform and steady \citep{anderson1990modern}, blast waves possess a decaying flow field behind them as well \citep{deweyAirVelocityBlast1964, murphyPIVSpacetimeResolution2010a}. A self-similar solution was proposed by \cite{taylorFormationBlastWave1950,taylorFormationBlastWave1950a}, \cite{von1955blast}, and \cite{sedov1946propagation} independently, with the blast wave speed and position being the similarity parameters. G. I. Taylor employed this solution to accurately estimate and disclose the classified energy of the atomic bomb during the Trinity nuclear test, using only publicly available photographs of the explosion \citep{taylorFormationBlastWave1950a}. This calculation led to difficulties with the U.S. government, which Taylor described as "much embarrassment." These details are documented in his biography, authored by his esteemed student G. Batchelor \citep{batchelor1996life}. But this similarity solution is suitable only for very strong explosions with high energy deposition and a stronger associated blast wave with higher Mach numbers. Later, this inspired others to pursue a solution for these waves across different strength regimes, involving clever approximations. Firstly, it is anticipated that, even in cases of intense blasts with high energies, the ongoing decay ultimately results in an asymptotic transition of the solutions from the strong shock regime to the acoustic regime, and scaling laws can be derived for both of these limiting conditions \citep{weiNewBlastWave2021a, diazBlastWaveKinematics2022a}. The intermediate phase of moderate blast strength poses a significant challenge, and self-similar solutions are not valid. \cite{sakuraiPropagationSphericalShock1956} introduced a perturbation solution assuming an approximate linear velocity profile, \cite{oshima1962blast} proposed a quasi self-similar solution, and \cite{bachAnalyticalSolutionBlast1970} employed a power-law density profile combined with the conserved energy integral. A more comprehensive overview of these approaches can also be found in the monograph by \cite{lee2016gas} and the review article by \cite{korobeinikovGasDynamicsExplosions1971}. Several studies even consider an equivalent source to mimic an explosion, for instance, \cite{lingAsymptoticScalingLaws2018a} presented a solution for the blast wave generated by a sudden release of a sphere of compressed gas, and \cite{radulescuTransientStartSupersonic2007} modeled an unsteady shock-supersonic jet system with a hypersonic virtual source.

We are particularly interested in the open-ended shock tube configuration where the shock waves or blast waves exit from the open end, accompanied by unsteady high-speed airflow emanating from a nozzle. This flow illustrates an energetic starting jet typically led by a compressible vortex ring (CVR) \citep{fernandezCompressibleStartingJet2017}. The impulsive action of this flow finds application in propulsion systems such as rocket nozzles and secondary maneuvering devices, such as cold gas thrusters, popular in reusable rockets. Such a flow also appears in the muzzle or open end of a barrel associated with exhaust ports or firearm-like devices, with the ejection of high-speed gases and projectiles \citep{merlenSimilarityAsymptoticAnalysis1991}. Understanding this flow is also necessary to understand the acoustic noise characteristics or screeching associated with the compressible jets \citep{suzukiShockLeakageUnsteady2003, fernandezCompressibleStartingJet2017}. The interaction of a steep impulsive wave in such flows also simulates the reflection of compression waves caused by high-speed trains approaching the tunnel entrance \citep{howeReflectionTransmissionCompression2005}. Accurate characterization of the airflow and efflux from the tunnel is essential to ensure the safety of the surrounding infrastructure. Volcanic eruptions also involve the rapid release of hot gases \citep{kiefferLaboratoryStudiesVolcanic1984}, during which volcanic vortex rings have been observed.

In the open-ended shock tube, the planar shock and associated flow evolution at the tube opening can be decomposed into the following major segments: (1) diffraction of the shock wave over the edge, (2) ejection of the trailing high-speed flow into a suddenly expanded cross-section. Earlier studies involved the interaction of plane shock waves with a convex wedge or corner with different angles \citep{skewsPerturbedRegionDiffracting1967, sunFormationSecondaryShock1997, sunVorticityProductionShock2003}. The shock when passing over such a corner with a sudden change in wall direction involves distortion into a curved shape, known as shock diffraction. This event is followed by the separation of the trailing gas flow around the corner, since it cannot follow the abrupt deflection of the wall and lead to the formation of a spiral vortex, which is well studied in the literature \citep{skewsPerturbedRegionDiffracting1967}. These studies also estimated the circulation production associated with such vortices \citep{sunVorticityProductionShock2003}. This isolated interaction of a shock wave with a corner, as will be discussed later, can be extended to the shock tube opening, which resembles a corner with right-angled geometry for blunt shock tubes or other specific angles pertaining to a beveled geometry. These effects culminate from all around the edge, leading to the formation of a curved shock front approaching an overall spherical shape far away from the opening. The separated vortex from the corner, when considered all around the shock tube, forms a vortex ring. This shock-induced starting jet and CVR have been of special interest in the past \citep{elderExperimentalStudyFormation1952, bairdSupersonicVortexRings1987, brouillettePropagationInteractionShockgenerated1997a, arakeriVortexRingFormation2004, fernandezCompressibleStartingJet2017, zhangEvolutionInitialFlow2019, rezayhaghdoostDynamicEvolutionTransient2020, poudelCharacteristicsShockTube2021}. The geometry and evolution of these CVRs significantly depend on the cross-section of the shock tube, i.e., the shape of the opening \citep{zare-behtashGlobalVisualizationQuantification2009, zare-behtashExperimentalInvestigationsCompressible2008}. In the early stages, they conform to the exact shape of the tube opening; however, the viscous effects resolve the portions of these vortex loops with large curvatures \citep{ghasemiCurvatureinducedDeformationsVortex2019,ghasemiViscousDiffusionEffects2018}. If the opening shape has singularities, for instance, a vertex in the case of any polygonal shape, this leads to the formation of dispersive kinks (waves) in the vortex loop. The dispersive kinks even lead to the swapping of the edge centers with the corners, often termed as corner inversion \citep{bauerEffectShockTunnel2023}. \cite{zare-behtashShockWaveinducedVortex2010} pushed it to the extreme by using eye-shaped nozzles to determine the role of singularities on CVR, and \cite{zamanSpreadingCharacteristicsCompressible1999} considered starting jets from even more complicated opening geometries. For an elliptical (or rectangular) opening, the CVR retains an elliptical (or rectangular) shape initially and exhibits unsteady evolution. In this process, different segments of the loop possess varying translational velocities, and there is a swapping of the major and minor axes of the evolving CVR in an oscillatory manner - an occurrence commonly referred to as axis switching \citep{zamanAxisSwitchingSpreading1996}. In the case of more complex openings, for instance, a rectangular feature which is of interest in the present study, this oscillatory motion becomes significantly more intricate \citep{grinsteinSelfinducedVortexRing1995, zamanAxisSwitchingSpreading1996, grinsteinVortexDynamicsEntrainment2001, koroteevaNumericalExperimentalStudy2016, ghasemiViscousDiffusionEffects2018, ghasemiCurvatureinducedDeformationsVortex2019}. In these investigations, Schlieren imaging facilitated the examination of the intricate flow feature resulting from compressible effects in the CVR and supersonic starting jet, including barrel shock, Mach stem, embedded shock, shear layers, slip stream, and counter-rotating vortices. Complementary numerical simulations provided further insight into the formation of these structures, which were often eluded by the limitations of experimental methodologies. Numerous experimental investigations utilized Particle Image Velocimetry (PIV) to quantitatively evaluate the flow dynamics, including velocity and vorticity distributions for the CVR \citep{arakeriVortexRingFormation2004, doraRoleSlipstreamInstability2014a,xiangCirculationProductionModel2023}. This approach enabled the characterization of CVR in terms of core features, circulation, the mechanism of the slip stream instabilities, and counter-rotating vortices. Several attempts have been made to model this compressible starting flow in terms of circulation production \citep{xiangCirculationProductionModel2023} and optimal formation number \citep{mohseni2001optimal} representing a dimensionless time or mass flux associated with CVR \citep{fernandezCompressibleStartingJet2017}; however, a comprehensive and reliable mechanism has not yet been established. Compounding this issue, the ejection of the shock wave from the open end and the associated interaction of incident waves lead to the formation of internally reflected waves. These crossing waves significantly alter the transient gas ejection dynamics at the opening. Additionally, the state of the flow behind the incoming shock, i.e., subsonic or supersonic, exerts a significant effect on the development of internal reflected characteristics and the external flow structures beyond the exit plane. \cite{rudingerImprovedWaveDiagram1955,rudingerReflectionShockWaves1955,rudingerReflectionPressureWaves1957,rudingerNonsteadyDischargeSubcritical1961} made notable efforts to model the reflection of waves associated with subcritical flows - including shock waves, compression waves, and rarefaction waves - at the open end of a shock tube, by imposing a boundary condition deduced from acoustic considerations, which we'll discuss later. However, the influence of these opening effects on gas ejection and transient jet formation remains insufficiently explored. Additionally, the existing literature primarily addresses incident shock waves traveling at constant speed with steady upstream flow. There is a significant gap in research concerning decaying incident flows and blast waves, and their evolution in open-ended shock tubes. Only a handful of investigations considered an unsteady flow at the open end, such as an oscillatory flow through an acoustic wave guide \citep{martinezdelrioGenerationVorticityOpen2023} and detonation-driven flow during blow-down of a pulse detonation combustor \citep{haghdoostHighSpeedSchlierenParticle2020}. As will be seen in this study, unsteady incident flow has significant implications on the evolution of outflow and associated flow structures, which consistently differ from their steady flow counterparts. 

In the present study, we investigate the propagation of a blast wave passing through the shock tube, its interaction with the tube opening, and the induced flow at the outlet. Such a system is expected to generate a controlled flow environment with a decaying flow profile emulating relevant scenarios. There are only a few alternate configurations that can generate such a flow in a controlled fashion. A diaphragm-based shock tube with a small driver section and/or a driver gas with a high speed of sound enables a similar flow configuration, where the reflected expansion waves catch up with the shock wave, leading to its decay. \cite{haselbacherOpenEndedShockTube2007} examined how the placement of the diaphragm influences flow behavior at the open end of shock tubes, with particular focus on the contact discontinuity and expansion fan. \cite{tasissaFormationFriedlanderWaves2016} developed a Friedlander wave \citep{friedlanderDiffractionSoundPulses1946}, or an approximate equivalent of a blast wave, resulting from a shock wave interacting with the expansion fan. They presented a solution based on the method of characteristics. The unsteady flow of interest is also synonymous with the transient discharge of a gas from a finite reservoir with varying stagnation properties \citep{orescaninExhaustUnderexpandedJets2010}.

The proposed system, involving a wire explosion-induced blast wave within a shock tube, generates an impulsive flow at the open end \citep{sembianPlaneShockWave2016c,livertsMitigationExplodingwiregeneratedBlastwaves2015}. A carefully designed converging conduit can even generate a supersonic jet at the opening by blast wave focusing \citep{apazidisSupersonicJetBlast2021a}. The currently used open-ended shock tube, featuring a rectangular duct and a copper wire as the explosive charge, has been strategically implemented earlier to study the atomization of a single drop of different liquids \citep{sharmaShockInducedAerobreakup2021c,sharmaShockinducedAtomisationLiquid2023a,sharmaDepthDefocusTechnique2023a,sharmaAdvancesDropletAerobreakup2022,chandraShockinducedAerobreakupPolymeric2023a,chandraElasticityAffectsShockinduced2024}, flames associated with a burning fuel droplet \citep{vadlamudiInsightsSpatiotemporalDynamics2024, vadlamudiEffectBlastWave2025}, and interaction with premixed and non-premixed combustible gas flames \citep{aravindResponseNonpremixedJet2025,aravindResponsePremixedJet2025}. Such interactions have been monumental in isolating and emulating the extreme flow fields relevant in various practical flow settings and gaining fundamental insights into complicated flow physics. On one end of the shock tube, the explosion creates a curved blast wave that passes through the rectangular tube cavity and is focused into a planar wave front. This blast wave then interacts with the opening, and there are reflected and transmitted waves. The efflux encompasses an unsteady starting jet with a leading asymmetric compressible vortex with embedded transient features such as shock wave shedding and smaller vortices. The experimental setup is outlined in Section~\ref{sec:Experiments}. Subsequently, a comprehensive analysis of the results is presented in Section~\ref{sec:Results}, focusing on blast wave dynamics, CVR, transient features (such as shear layer, expansion wave, and secondary shocks), and the induced flow field at the tube opening, achieved through the application of the method of characteristics employing alternative boundary conditions at the tube opening. This investigation seeks to facilitate a comprehensive understanding of the complex spatiotemporal dynamics of multiple features associated with highly transient, blast-driven flow phenomena originating from an open-ended shock tube.

\section{Experimental Setup}
\label{sec:Experiments}

\begin{figure}
  \centerline{\includegraphics[width=1\linewidth]{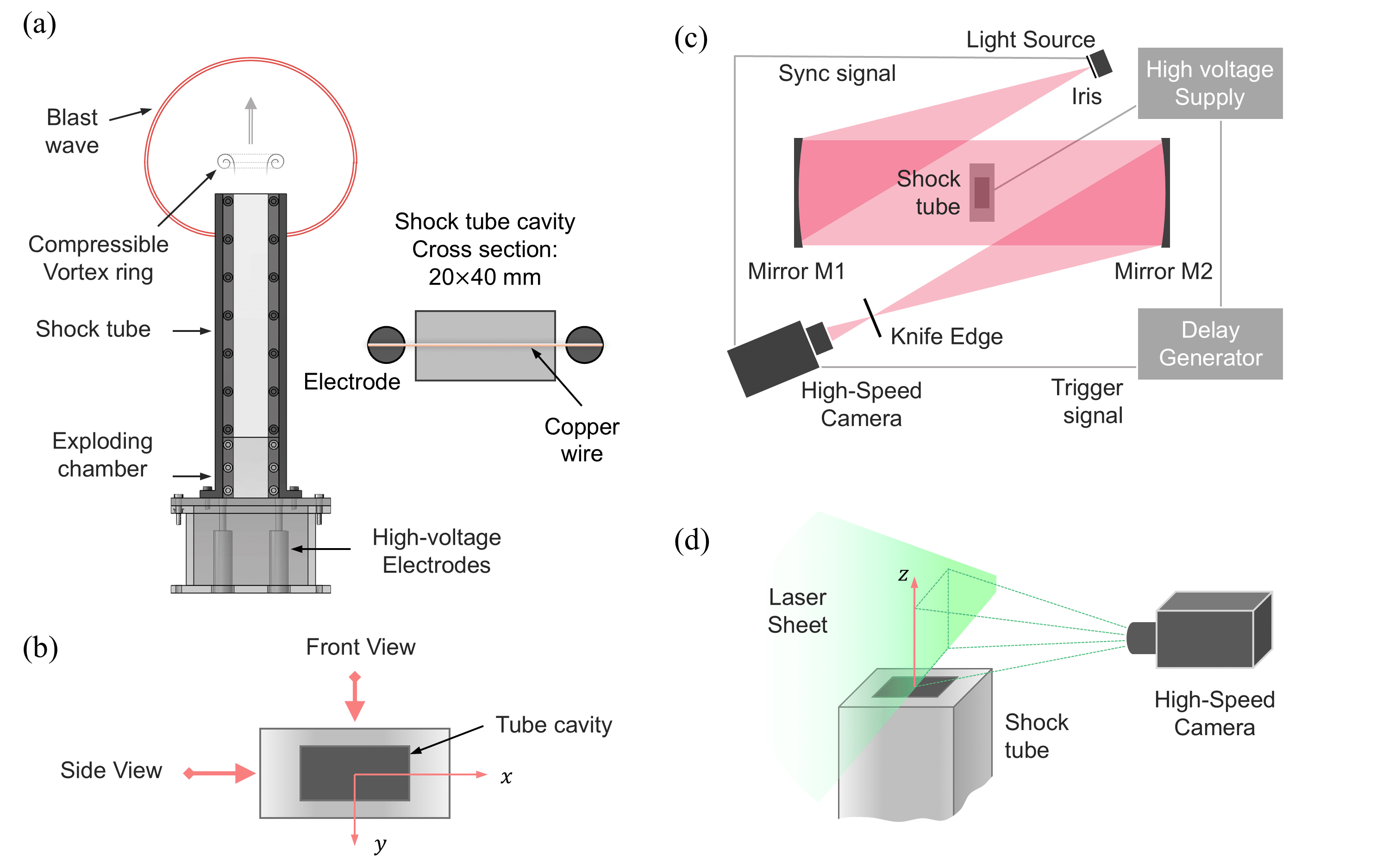}}
  \caption{Schematic illustrating experimental setup (a) Wire explosion based shock tube and the rectangular cross-section of the cavity (b) Two primary orthogonal views for flow visualisation - Front View and Side view (c) Schlieren imaging setup to visualise the shock wave and the density gradients in the compressible vortex flow (d) Particle image velocimetry (PIV) setup to assess flow fields using tracers and a planar laser sheet illuminating one of the principal planes.}
\label{fig:setup}
\end{figure}

The present study deploys a wire-explosion-based approach to generate a controlled blast and the associated blast wave. A pulsed high-voltage system (2kJ pulse power, Zeonics Systech India, Z/46/12) is used to deposit electrical energy into a thin copper wire (35 SWG, bare copper) placed across electrodes for a very short time duration ($\sim 10^0\mu s$), as depicted in figure~\ref{fig:setup}a. This leads to the rapid heating and vaporization of the wire, creating an explosion accompanied by a cylindrical blast wave. This is placed at the bottom of a shock tube with a cavity having a rectangular cross-section ($20mm \times 40mm \times 325mm$). We define the edge dimensions of the rectangular cross-section as $D_1 = 40mm$, $D_2 = 20mm$, and the tube length as $L = 325mm$.  This asymmetry of the cross-section is necessary to focus the cylindrical blast generated from a line source into a planar one as it travels from the blast source toward the shock tube opening. This planar blast wave exits the tube and takes up a three-dimensional shape due to effects like diffraction. The accompanying unsteady flow dynamics at the tube opening are the focus of this paper. More details of this experimental setup can also be found in the previous works where atomization events in such extreme flow conditions were investigated \citep{sharmaShockInducedAerobreakup2021c, sharmaShockinducedAtomisationLiquid2023a}. 

A high-speed Schlieren and Mie-scattering imaging system is deployed to assess this transient phenomenon. A digital delay generator (BNC 575) is employed to trigger the imaging and shock tube system simultaneously. The shock Mach numbers ($Ma_s$) are controlled by adjusting the charging voltage ($V_c$) associated with the capacitor, corresponding to the high voltage supply. Experiments are carried out by varying the charging voltages ($V_c = 6kV-10kV$) with corresponding shock Mach numbers $Ma_s=1.2-1.8$, as measured and illustrated in the subsequent sections. The asymmetry of the system is acknowledged by visualizing the flow in the two principal planes identified as the front view (FV, i.e., $xz$-plane) and the side view (SV, i.e., $yz$-plane) as illustrated in figure~\ref{fig:setup}b. The shock tube is rotated along the $z$-axis within the flow visualization setups to capture the two orthogonal views. The shock tube is constructed from polycarbonate panels, allowing for direct optical access to observe internal shock wave dynamics via front-view visualization.

A schlieren imaging setup is illustrated in figure~\ref{fig:setup}c. It consists of two concave mirrors ($145mm$ diameter and $1.5m$ focal length) to generate a parallel beam of light. A high-speed pulse diode laser source ($640 nm$ wavelength, Cavitar Cavilux smart UHS, 400 W power), along with an variable opening iris aperture (Holmarc SSID-25) and a lens apparatus generates a diverging beam of light, which is placed at the focus of the first mirror M1 generating a parallel beam of light passing through the region of interest, i.e., the shock tube opening, as depicted in figure~\ref{fig:setup}b. The second mirror M2 is placed on the other side to capture this beam and converge it to approximately a point (focus), where a knife-edge is placed to partially block the light. As the flow is along the vertical axis with a significant density gradient in the vertical direction, the knife-edge was placed horizontally. A high-speed camera (Photron SA5) is then used to capture the images at high acquisition rates, ranging from $20000-40000 Hz$. The images were captured at a spatial resolution of $0.294 mm/$pixel. The camera and the pulsed light source were synchronized during the recording operation.

The setup for the Mie-scattering flow visualization is illustrated in figure~\ref{fig:setup}d. Olive oil droplets ($\sim1\mu m$ diameter) were used as tracer particles and were introduced to the shock tube cavity before initiating the blast. A lid was used to ensure that the cavity was uniformly seeded, and this lid was removed before the system was triggered. A high-speed dual-pulsed Nd:YLF laser (wavelength $532 nm$, pulse energy $30 mJ$ per pulse) is deployed as an illumination source. The laser beam is converted into a thin sheet $(1 mm)$ using sheet optics, aligned with a principal plane to illuminate and visualize the induced flow at the tube opening. A Photron Mini UX-100 high-speed camera was used to capture the images at $2400Hz$ with a double pulse at an interval of $5\mu s$. A field of view of $1280\times 1024$ pixels was obtained with a spatial resolution of $56.75 \mu m/$pixel. The raw images were processed using the Davis 8.4 software (Lavision GmbH) and PIVlab \citep{Thielicke_2021}. The processing involved a cross-correlation multi-pass algorithm with a decreasing window size from $64 \times 64$ pixels to $32\times 32$ pixels (three passes) with $50\%$ overlap to obtain the velocity vector field. This ensured a vector spacing of $\sim 0.91 mm$ in both directions ($xz$ or $yz$). A bi-cubic reconstruction interpolation was implemented, along with a multi-pass median filter in the final pass to eliminate noise. As the acquisition rate is inadequate considering the extreme speed of events, $20-30$ runs were taken in a staggered fashion with a predetermined lag, and an ensemble of all the runs was compiled to determine the centerline axial velocity, presented later. Furthermore, PIV implementation with this particular setup entailed an additional challenge, which is currently unavoidable. The explosion of the copper wire generates an intense flash of light that lasts for $\sim 1 ms$, which adversely affects PIV measurements. This intense flash illuminates the whole volume, along with all the tracers contained within, in the early stages of vortex evolution. This obstructs the intended planar measurement with PIV, and the associated data is unreliable and disregarded in the present study. A band-pass filter associated with the wavelength of the monochromatic laser source did not help, since the severe poly-chromatic flash from the blast also contained this segment of the light spectrum. This effect was even more pronounced for blasts with higher energies, having a brighter flash that lasted longer. The obtained visualization and flow fields are discussed in detail in the subsequent sections.

\section{Results and Discussion}
\label{sec:Results}

\subsection{Global assessment of the phenomenon}
\begin{figure}
  \centerline{\includegraphics[width=1\linewidth]{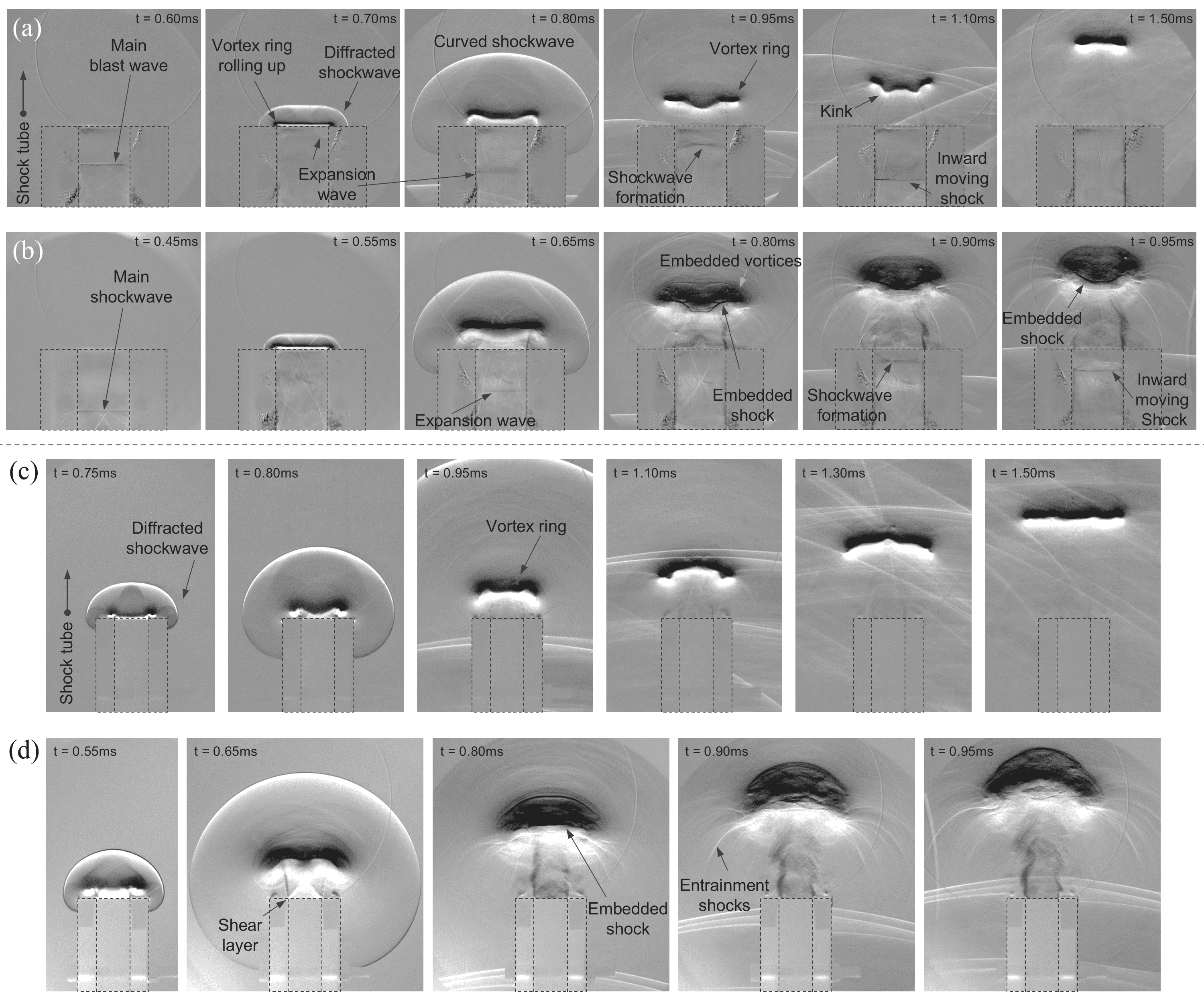}}
  \caption{Schlieren imaging of the flow evolution near the shock tube exit illustrating primary blast wave, diffracted shockwave, expansion wave, compressible vortex ring (CVR) and reverse shock wave from the front view at (a) lower blast energy at $V_c=6kV$ (b) higher blast energy at $V_c=10kV$ and the side view at (c) $V_c=6kV$ (d) $V_c=10kV$.}
\label{fig:schlieren}
\end{figure}

A blast is initiated at the bottom of the shock tube by exploding a wire. The associated cylindrical blast wave originating from the closed end of the tube is focused within the shock tube cavity and has a planar structure as it reaches the opening. This flow feature appears as a line inside the shock tube in the Schlieren images, as illustrated in figure~\ref{fig:schlieren}a,b, representing a density discontinuity. Accounting for the unsteady nature of this shock wave, the propagation speed and the accompanying flow field are expected to vary with time, showing a decaying behavior as will be evident from the measurements and analysis in the subsequent sections. As the rectangular tube opening imposed asymmetry, the flow dynamics through schlieren were captured from both the principal axes, namely FV and SV, represented in figure~\ref{fig:schlieren}a,b, and figure~\ref{fig:schlieren}c,d, respectively.

The shock wave (or blast wave), as it reaches the exit plane, diffracts at the corner or inner lip of the shock tube opening. Locally, this phenomenon is equivalent to the shock diffraction over a convex wedge with a 90-degree turn. This sudden turning of shock is accompanied by morphing into a curved shape and an expansion fan propagating upstream. From the perspective of the tube opening, this is equivalent to the sudden expansion of the duct, abruptly to an infinite area \citep{peacePropagationDecayingPlanar2018a}. The diffraction-induced curvature from the edges eventually reaches the shock tube centerline, and the overall shock shape then conforms to approximately an ellipsoidal shape \citep{newmanPeakOverpressureField2010}, corresponding to the asymmetric effects arising from the rectangular opening. For an axisymmetric source, i.e., a circular opening, the shock is expected to take a spherical shape after a short lag period after it exits the shock tube. For the non-axisymmetric source, with the present rectangular geometry, the ellipsoid is expected to evolve into a spherical shape at very large radial distances from the source, where the point source approximation is valid \citep{chiuBlastWavesAsymmetrical1977}. This equivalent point source is expected to be at or very close to the exit plane at the shock tube end, as will be seen later.
The blast wave inside the tube is accompanied by a decaying flow field. Neglecting the boundary layer effects or viscosity, which is relevant for shock tubes with very small cross-section \citep{janardhanrajInsightsShockwaveAttenuation2021a, mirshekariOnedimensionalModelMicroscale2009}, this can be assumed to be a one-dimensional system with minimal flow variation across the cross-section. 

\begin{figure}
  \centerline{\includegraphics[width=1\linewidth]{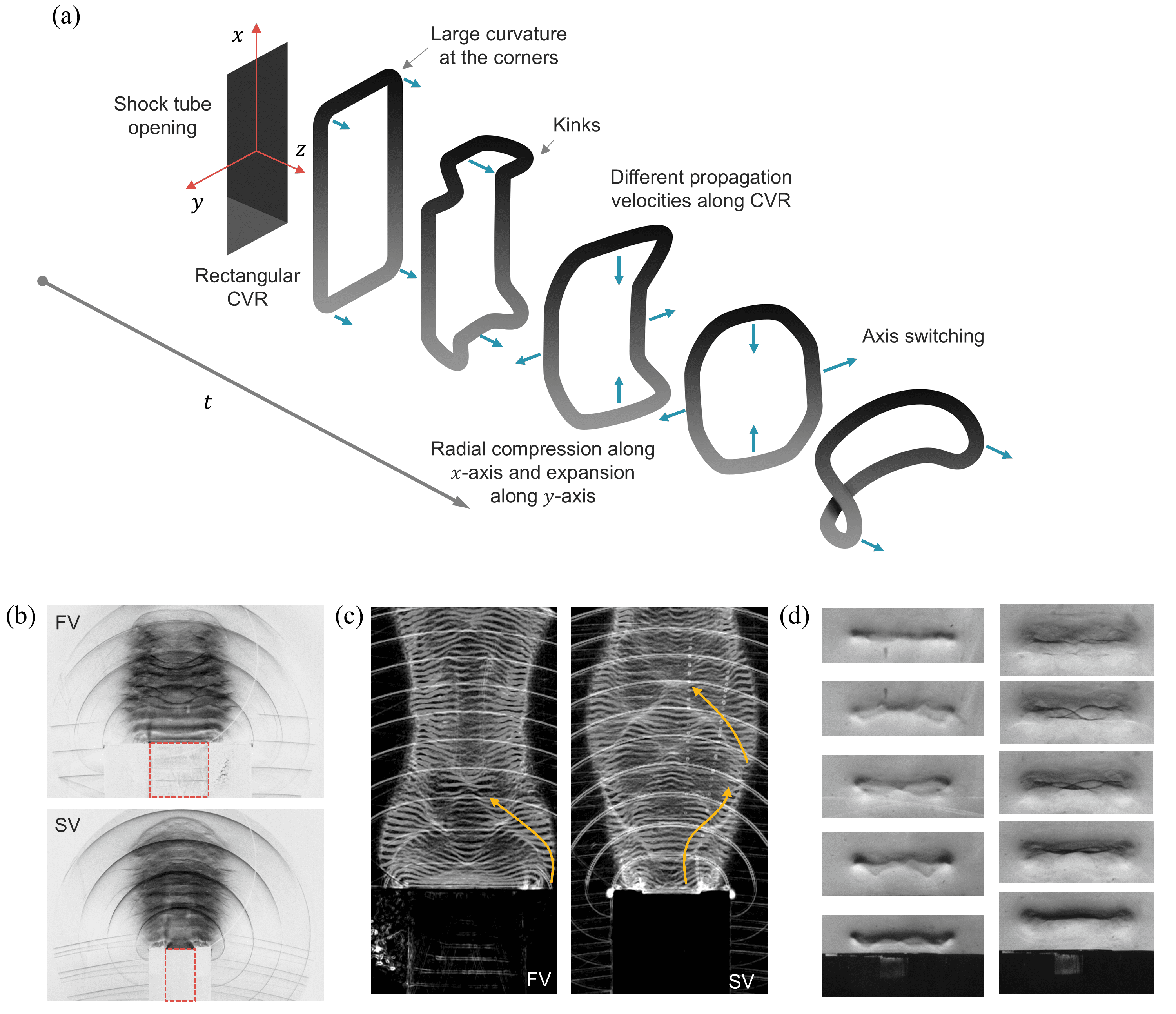}}
  \caption{(a) Schematic illustrating the asymmetric evolution of the compressible vortex ring (CVR) displaying kinks, unsteady deformations, and axis switching. (b) Schlieren images from the front view (FV) and side view (SV) superimposed over a time period illustrating an asymmetric flow. The red dotted box illustrates the shock tube cavity. Top: FV and Bottom: SV for $V_c=6kV$ (c) Superimposition of processed schlieren images (gradient operation), showing the trajectory of the kinks over the unstable CVR through yellow arrows. Left: FV and Right: SV for $V_c=6kV$ (d) Diagonal view of the vortex ring. Left: $V_c=6kV$ with slender vortex and Right: $V_c=10kV$ with curved embedded shock.}
\label{fig:phenomenon}
\end{figure}

As the blast wave reaches the opening, the sudden expansion offers an additional degree of freedom. As the blast wave diffracts, the accompanying flow rolls up at the inner lips of the tube opening, forming a CVR. The initial shape of the CVR aligns with the rectangular geometry.  The blast propels ahead; however, the CVR carrying most of the induced flow follows at a lower convective speed. The schematic illustrating the CVR evolution is presented in figure~\ref{fig:phenomenon}a. The four corners associated with the rectangular lip give rise to singular kinks in the compressible vortex loop, which are resolved due to viscous effects and appear as waves over the loop. The part of the loop originating from the shorter edge of the rectangle propels faster since this segment has a lower radius of curvature, following the Biot-Savart law. The same logic is applicable to the loop segments at the corner of the rectangle, which possess extremely large curvatures in the beginning and hence translate faster than the rest of the CVR initially. The asymmetry also causes self-induced deformation in the CVR and causes large-scale loop oscillations \citep{ghasemiViscousDiffusionEffects2018, ghasemiCurvatureinducedDeformationsVortex2019}. This also leads to axis-switching and, in extreme cases, has been considered previously in the literature, even CVR splitting into multiple loops \citep{grinsteinVortexDynamicsEntrainment2001}.  

\begin{figure}
  \centerline{\includegraphics[width=1\linewidth]{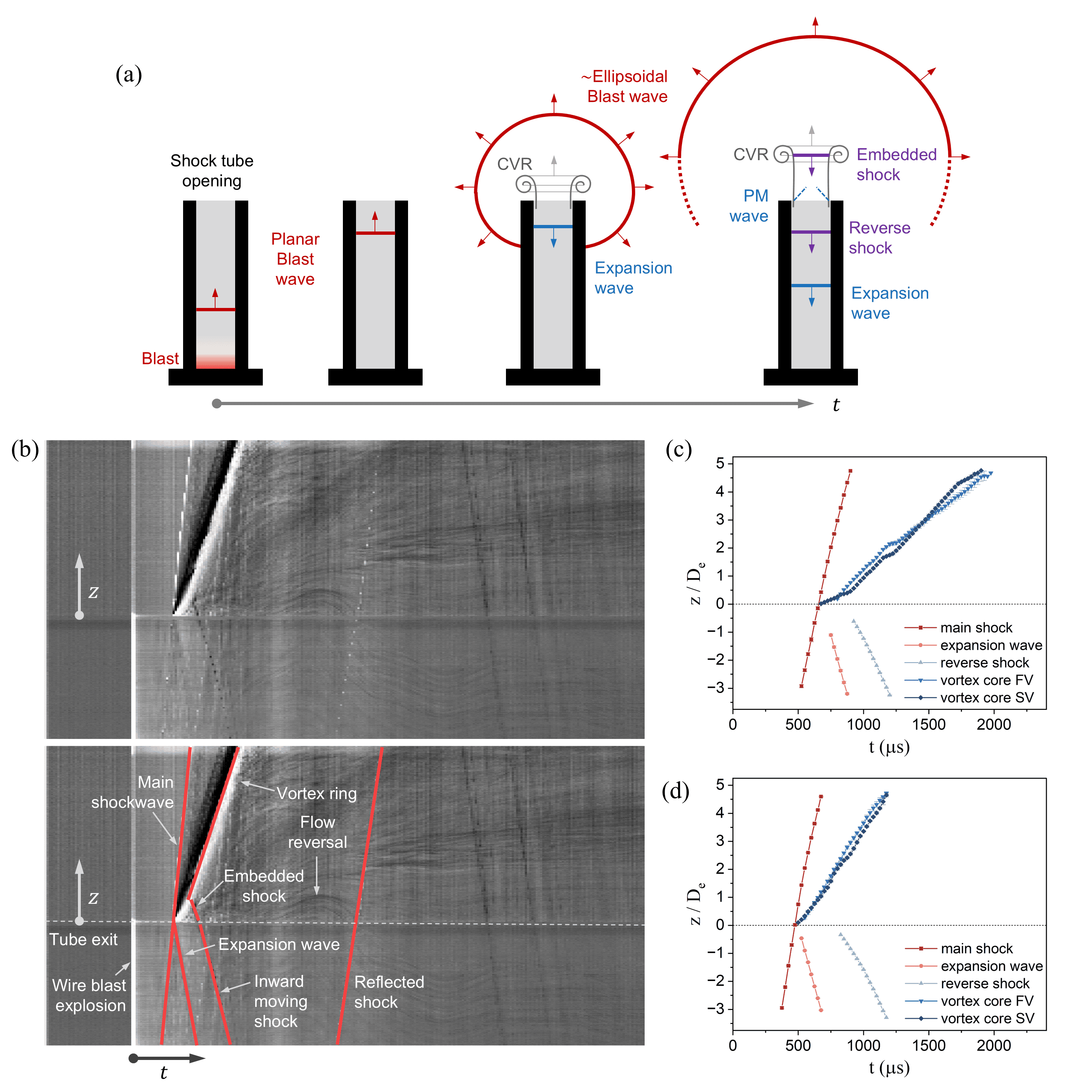}}
  \caption{(a) Schematic illustrating the blast wave transmission across the tube exit (b) Line schlieren: Temporal evolution of the centerline intensity of the Schlieren imaging of the front view, plotted as a contour diagram for blast energy $V_c=8kV$, illustrating evolution of the significant flow features. $z-t$ diagram extracting spatiotemporal information of the significant flow features for blast energies (c) $V_c=6kV$ (d) $V_c=10kV$, where $z/D_e=0$ marks the tube exit. CVR: compressible vortex ring; PM: Prandtl-Meyer wave}
\label{fig:x-t}
\end{figure}

Superimposing the subsequent schlieren images (through an averaging operation on the image stack) results in an overexposed composite image, as shown in Figure~\ref{fig:phenomenon}b for both principal directions. The images clearly demonstrate an asymmetric evolution of the flow when observed from the FV and SV perspectives. To observe the motion of undulations on the vortex loop originating due to the rectangular tube geometry, a gradient operator is imposed over the schlieren images, and processed consecutive images are further superimposed (through a standard deviation operation on the image stack) and illustrated in figure~\ref{fig:phenomenon}c. The kinks originate from the corner and move over the loop as a wave towards the mid-plane bisecting the longer edge of the shock tube ($yz$-plane). In other words, they appear to move towards the center when viewed from the FV as illustrated through arrows in figure~\ref{fig:phenomenon}c. These waves from all the corners interact and pass each other as can be seen in figure~\ref{fig:phenomenon}. The shock tube was placed diagonally in the collimated light beam, presenting an intermediate between FV and SV for schlieren imaging. This view offers an insight into the three-dimensional evolution of this wavy vortex loop as shown in figure~\ref{fig:phenomenon}d. Apart from these oscillations of the unstable CVR, we observe other secondary features, such as shear layers, embedded shocks, and entrainment shocks. At lower blast intensities, the CVR demonstrates a thin core, as observed in the Schlieren images in figure~\ref{fig:schlieren}a,c. As the blast intensity is increased, the CVR core gets thicker and more turbulent, as apparent from the small-scale density fluctuations captured in the experiments in figure~\ref{fig:schlieren}b,d. We also see a shear layer originating from the tube exit, as part of the trailing jet behind the CVR. Furthermore, we observe an embedded shock lying within the CVR, spanning core to core and illustrating a shedding pattern. These move inward towards the tube opening and disappear as they weaken, causing a subsequent embedded shock to reappear within the CVR. These shocks are cup-shaped with a flat center, as shown in figure~\ref{fig:schlieren}b,d and~\ref{fig:phenomenon}d. This embedded shock topologically resembles a distorted barrel shock-Mach disc combination, typically apparent in the under-expanded supersonic starting jets with a steady supply. The current system involves a decaying supply from a finite-sized reservoir, which might contribute to the unsteady shedding. This phenomenon is further discussed in the subsequent sections.

From the perspective of the CVR core, in crude terms, the air has to travel around in the direction: CVR center line – forward – outer periphery – back to centerline. During this, the air from the surroundings is entrained as well, particularly contributing to the peripheral and trailing regions. In this process, especially with compressible effects, the flow locally depicts a transonic behavior, normalized by the presence of entrainment and embedded shocks \citep{bairdSupersonicVortexRings1987, brouillettePropagationInteractionShockgenerated1997a, rezayhaghdoostDynamicEvolutionTransient2020}. These shocks are visible in CVR associated with higher blast intensity and also illustrate unsteady characteristics in the current system, originating from the decaying source flux.

As the blast wave exits the tube, the expansion fan originating from the peripheral lip moves inwards. With time, they are superimposed to form a planar expansion wave that travels inside the tube, as visible in figure~\ref{fig:schlieren}a,b. This unsteady feature affects the effective mass flux from the shock tube, which contributes to the CVR formation. This wave, after a brief time period, is followed by a new shock wave that originates near the opening and travels inwards, and can be seen in figure~\ref{fig:schlieren}a,b. This reverse shock wave or 'reshock' is observed across all the blast intensities in the current study. The exact mechanism will be discussed subsequently; however, to summarize briefly, the efflux vacates the air inside the cavity and, when combined with the expansion fan and transonic flow, leads to wave steepening. Since the pressure at the tube exit eventually or almost immediately normalizes to ambient pressure (higher than the tube cavity), this system is synonymous with an inverted shock tube momentarily. This also leads to a reversed flow inside the tube, as visible in the Schlieren images as well, where the trailing jet is eventually pinched off, followed by the reversed motion of flow features.


The phenomenology and development of the aforementioned features are broadly illustrated in figure~\ref{fig:x-t}a, providing a global perspective. Depending on the regime of outflow, i.e., subsonic or supersonic, we see other secondary shock features in the CVR and Prandtl-Mayer expansion waves outside the tube shown in later stages of figure~\ref{fig:schlieren}b,d, and \ref{fig:x-t}a and will be thoroughly assessed subsequently. The intensity history from Schlieren serves as an important tool to track these features in space and time \citep{bagaiFlowVisualizationCompressible1993a}, i.e., to generate the standard $x-t$ representations popular in such studies: a $z-t$ diagram in our case, based on the present coordinate system. The centerline intensity at each time frame is stacked to create a graphic, as illustrated in figure~\ref{fig:x-t}b, which is also known as a line schlieren \citep{haghdoostHighSpeedSchlierenParticle2020}, where various flow features are marked. When tracked, the quantitative information of the trajectory in the $z-t$ diagrams, shown in figure~\ref{fig:x-t}c, is used to determine shock arrival time and Mach numbers. Also, the trajectories of the primary and secondary shock features and CVR evolution can be extracted. The information inferred from these images is discussed in detail in the subsequent sections.

\subsection{Blast wave characteristics}

Blast wave dynamics involve two prominent stages in the current configuration: evolution inside and outside the shock tube. The focused blast wave inside the tube can be considered a one-dimensional system having planar geometry, assuming a minimal influence of the boundary layers near the cavity walls. Using the methodology presented by \citet{bachAnalyticalSolutionBlast1970, lee2016gas} for blast waves with moderate strengths (which is valid for the Ms range of current experiments, i.e., $1.02 < M_s < 1.6$), a power-law density profile can be assumed behind the blast wave, and the velocity profile can be obtained from the mass conservation differential equation. An effective impulsive energy of the blast source is to be assumed and back calculated using the shock trajectory obtained from the experiments, as will be discussed soon.

Let us consider a one-dimensional coordinate system $(r,t)$ with a blasting source with energy $E_0$ initiated at $(0,0)$. Firstly, we define the non-dimensional parameters as follows:
\begin{equation}
\xi=\frac{r}{R_s\left(t\right)},\ \ \eta=\frac{{a_o}^2}{{{\dot{R}}_s}^2}=\frac{1}{{M_s}^2}
\end{equation}
\begin{equation}
\phi\left(\xi,\eta\right)=\frac{u(r,t)}{{\dot{R}}_s(t)},\ \ f\left(\xi,\eta\right)=\frac{p(r,t)}{\rho_0\dot{R}_s^2},\ \ \psi\left(\xi,\eta\right)=\frac{\rho(r,t)}{\rho_0} 
\end{equation}
where $R_s$ is the instantaneous shock radius, ${\dot{R}}_s$ is the shock speed, $u$ is the air (or gas) velocity, $\rho$ is the air density, $p$ is absolute pressure,  $a$ is the speed of sound and $M_s$ is the shock Mach number. The subscript $'0'$ denotes the pre-shock ambient conditions. The conservation of mass and momentum for the unsteady one-dimensional adiabatic motion of a perfect gas behind the expanding blast wave in terms of these normalized parameters is given as:
\begin{equation} \label{eq:mass}
\left(\phi-\xi\right)\frac{\partial\psi}{\partial\xi}+\psi\frac{\partial\phi}{\partial\xi}+j\phi\frac{\psi}{\xi}=2\theta\eta\frac{\partial\psi}{\partial\eta}\ 
\end{equation}

\begin{equation} \label{eq:momentum}
\left(\phi-\xi\right)\frac{\partial\phi}{\partial\xi}+\theta\phi+\frac{1}{\psi}\frac{\partial f}{\partial\xi}=2\theta\eta\frac{\partial\phi}{\partial\eta} 
\end{equation}

where $j$ is the geometry parameter such that $j=0, 1, 2$ and correspondingly $k_j = 1, 2\pi, 4\pi$ for planar, cylindrical, and spherical blast waves, respectively. $\theta$ represents the decay coefficient and is defined as
\begin{equation}  \label{eq:theta}
    \theta\left(\eta\right)=\frac{R_s{\ddot{R}}_s}{{\dot{R}}_s}
\end{equation}

The boundary conditions corresponding to the shock front $\xi = 1$ are obtained from Rankine-Hugoniot conditions

\begin{eqnarray}
    \phi\left(1,\eta\right)=\frac{2(1-\eta)}{\gamma+1},\ \ \ f\left(1,\eta\right)=\frac{2}{\gamma+1}-\frac{\gamma-1}{\gamma(\gamma+1)}\eta,\ \ \ 
\psi\left(1,\eta\right)=\frac{\gamma+1}{\gamma-1+2\eta}
\end{eqnarray}

\citet{bachAnalyticalSolutionBlast1970} assumed a power law for the density variation as follows
\begin{equation}
    \psi\left(\xi,\eta\right)=\psi(1,\eta)\xi^{q(\eta)}
\end{equation}
with the exponent $q$ establishing a decaying profile. At this stage, $q(\eta)$ and $\theta(\eta)$ are unknown. Two necessary conditions are enforced to determine these parameters, namely, the conservation of mass and energy bound within the control volume enclosed by the blast wave and blast origin. It signifies that the total mass and energy (originating from $E_0$) behind the blast wave is conserved and is redistributed as the system evolves in time. Imposing this condition generates the mass and energy integral equations \citep{lee2016gas}, setting up an integral approach to solving the problem, synonymous with integral methods in boundary layer theory. Readers can refer to \citet{bachAnalyticalSolutionBlast1970} and \cite{lee2016gas} for a detailed solution, which was implemented in this paper. In case of very strong Blast waves ($\eta \rightarrow0$), the constant coefficients $q$ and $\theta$ can be preemptively estimated from the blast geometry alone, and associated solutions are presented in Appendix~\ref{appA}. 

To reintroduce time, i.e., estimate the temporal variation of flow parameters, the shock trajectory is determined from ${\dot{R}}_s = dR_S/dt$, which can also be rewritten in terms of normalized parameters as 
\begin{equation}
    \frac{a_0t}{R_0}=-\frac{1}{2}\int_{0}^{\eta}\frac{y^\frac{1}{j+1}}{\theta\eta^{\frac{1}{2}}}d\eta
\end{equation}

where $y=\left(R_s/R_0\right)^{j+1}$ and $R_0$ is the characteristic explosion length that naturally appears while normalizing the energy integral and is defined as
\begin{equation} \label{eq:R_0}
    R_0 = \left(\frac{E_0}{k_j\gamma p_0}\right)^\frac{1}{j+1}
\end{equation}

\begin{figure}
  \centerline{\includegraphics[width=1\linewidth]{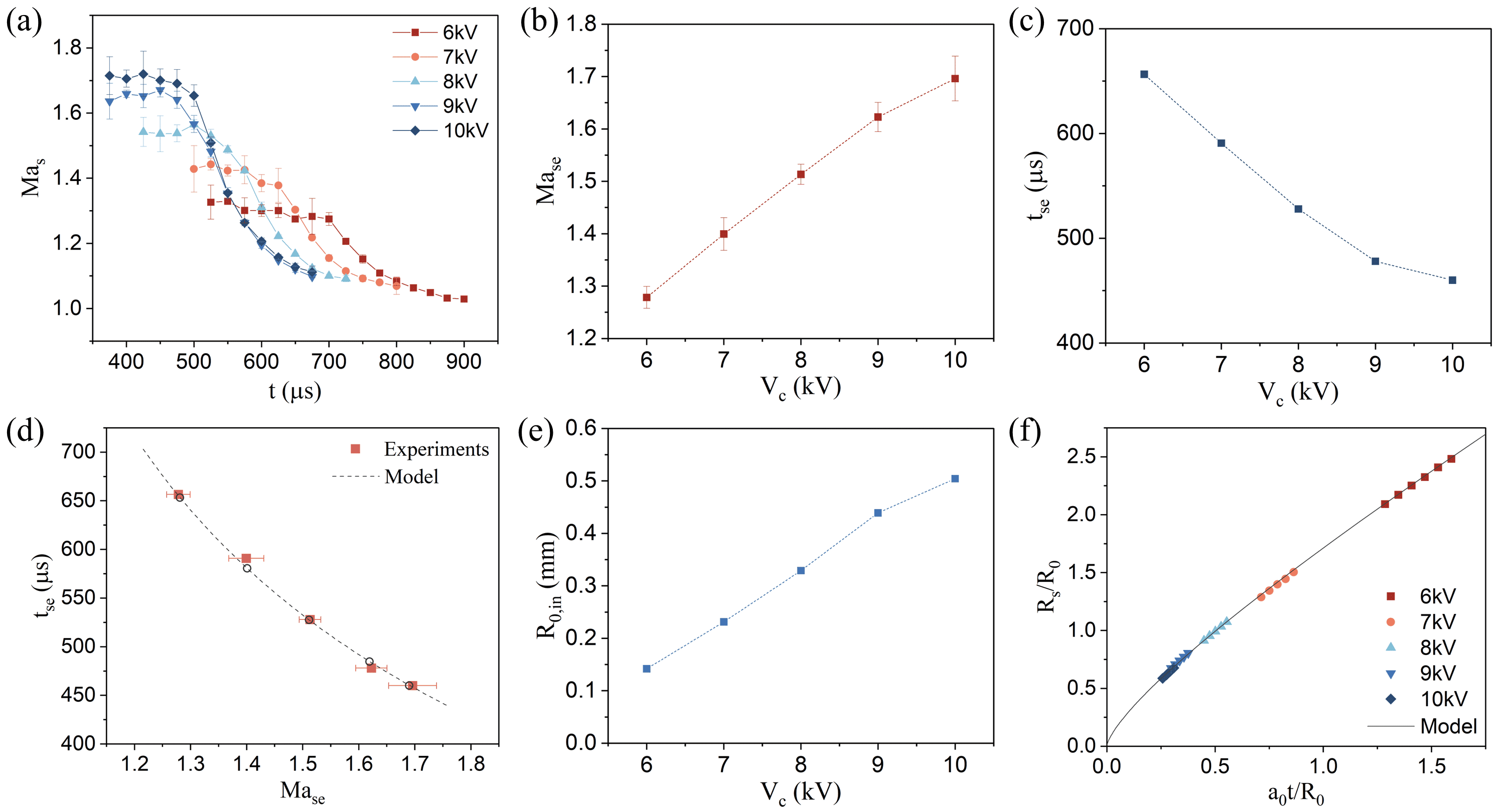}}
  \caption{(a) Time series of the blast wave velocity near the tube opening, illustrating significant decay outside the exit. Arrival properties of the blast wave at the tube exit at different charging voltages (b) Shock Mach number $Ma_{se}$  (c) Arrival time $t_{se}$. (d) Prediction of the arrival characteristics based on the model and (e) corresponding characteristic length $R_{0,in}$ values associated with blast energy from the system inside the shock tube. (f) Normalised blast wave trajectories inside the shock tube and comparison with the model.
}
\label{fig:shock_inside}
\end{figure}

As we cannot directly measure the blast energy in the present configuration \citep{chandraShockinducedAerobreakupPolymeric2023a,vadlamudiInsightsSpatiotemporalDynamics2024}, we estimate $R_0$ from the shock trajectory shown in figure~\ref{fig:shock_inside}a and use equation~\ref{eq:R_0} to determine $E_0$. The arrival properties of the blast wave at the opening of the shock tube are considered as a reference to estimate the blast model parameters. Shock mach number $Ma_{se}$ at the opening and arrival time $t_{se}$ is determined from the $z-t$ diagram (see. Figure~\ref{fig:x-t} and~\ref{fig:shock_inside}a) as illustrated in figure~\ref{fig:shock_inside}b,c.

\begin{figure}
  \centerline{\includegraphics[width=1\linewidth]{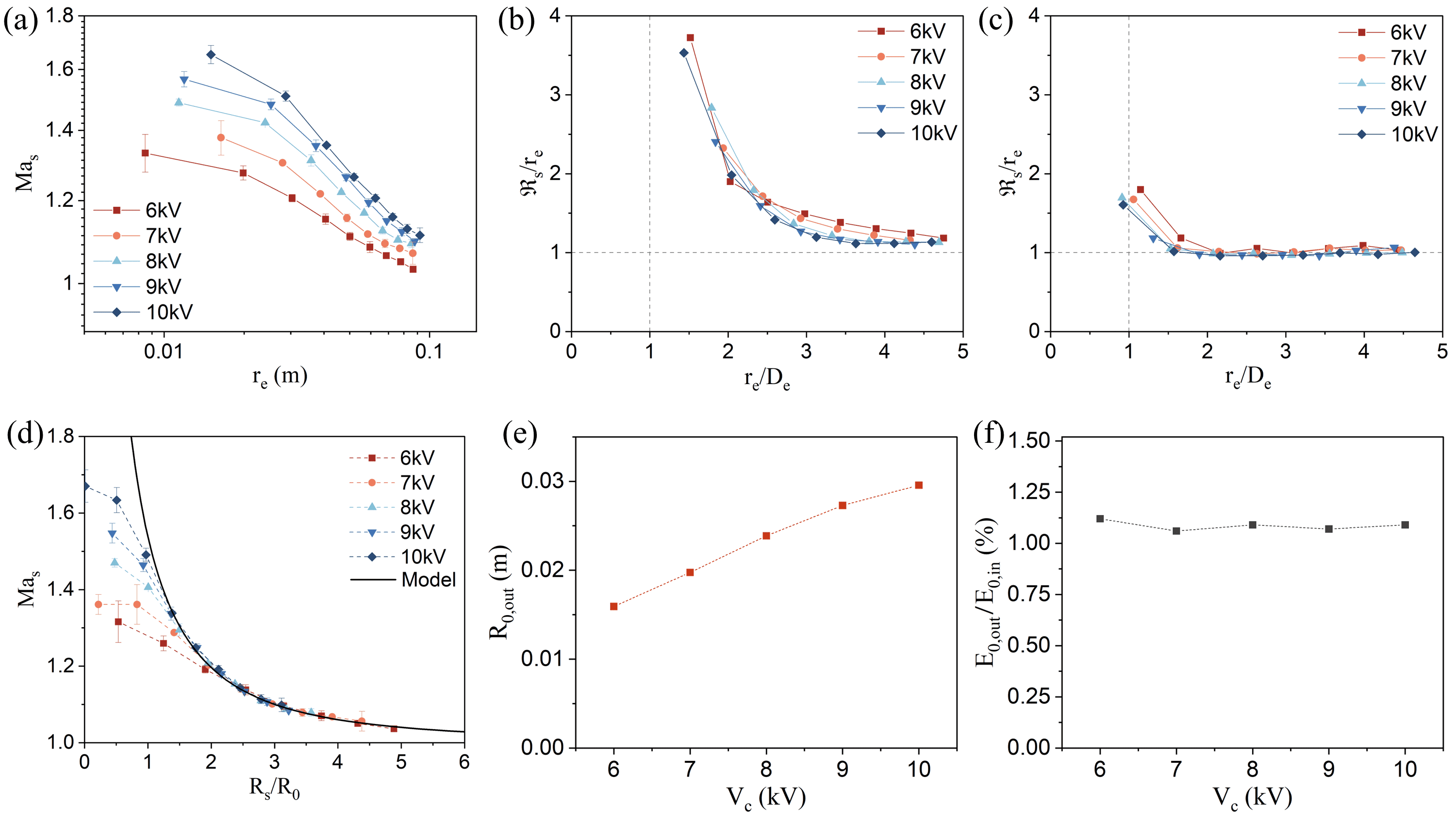}}
  \caption{(a) Blast wave velocity outside the shock tube exhibiting approximately a power law decay on a log scale. Evolution of the centerline curvature  $\Re_{s}$ of the decaying blast wave outside the tube, normalized with distance downstream the tube exit $r_e$ when viewed from (b) FV (c) SV. (d) Normalized blast wave evolution outside the shock tube with the tube exit as the center, and comparison with the model. (e) corresponding characteristic length $R_{0,out}$ values for the blast system outside the tube exit. (f) Ratio of the energies $E_{0,out}/E_{0,in}$ for the blast wave system inside and outside the shock tube. }
\label{fig:shock_outside}
\end{figure}

The tube exit is located at $r=L$, where $L = 325mm$ is the tube length. However, we vary $L$ as we iterate and estimate $Ma_{se}-t_{se}$ from the model using different $R_0$ until the shock arrival properties match the experiments. We find the match at an effective tube length of $r=L_{\rm eff}=355mm$. This offset is expected since the blast initially originates as a cylindrical wave, and because of the confinement, it eventually focuses into a planar wave. For the latter, the virtual source is located beyond the actual blast source, such that the opening is located at $r=L_{\rm eff}>L$. It is interesting that $L_{\rm eff}\approx L+D_2$. The comparison of experimental observations with the model is illustrated in figure~\ref{fig:shock_inside}d. The corresponding $R_0$ value for each case is displayed in figure~\ref{fig:shock_inside}e. It should be noted that these parameters associated with the model are only valid for the blast wave inside the shock tube. The normalized trajectory and its comparison with experiments in figure~\ref{fig:shock_inside}f shows that the model aligns well with the observations.

The blast trajectory outside the shock tube illustrates a different geometrical shape transitioning from a planar shape to ellipsoidal due to diffraction at the inner lip, as was shown in figure~\ref{fig:schlieren}. The additional degree of freedom entails higher decay rates \citep{peacePropagationDecayingPlanar2018a}, as illustrated through $Ma_s$ evolution in figure~\ref{fig:shock_inside}a. The slope associated with decay in figure~\ref{fig:shock_outside}a for a short span in space and time (log-scale) corresponds to $\theta$, with tube exit as the updated virtual center (for constant $\theta$ in equation~\ref{eq:theta}, ${\dot{R}}_s\sim R_s^\theta$ (see Appendix~\ref{appA}). To validate the position of the virtual center for the blast wave system outside the tube exit, we consider the radius of curvature $\Re_s$ of the blast wave at the centerline along two principal planes, measured from Schlieren images. For the blast position with tube exit as reference, we define a radial coordinate system with $r_e \equiv z$, the virtual source resides here if $\Re_s/r_e=1$. On $xz$-plane corresponding to FV, $\Re_s/r_e>1$ and barely approaches 1 for $r_e>D_2$ (see figure~\ref{fig:shock_outside}b). On $yz$-plane corresponding to SV, $\Re_s/r_e=1$ for $r_e>D_2$ (see figure~\ref{fig:shock_outside}c). This is indicative of the fact that the information of the presence of a sharp corner where the diffraction originates takes $r_e \approx D_2$ to reach the centerline. Since $D_2 < D_1$, $D_2$ appears as the representative scale, and we represent it as $D_e = D_2$ here onwards. Before this, the central part of the shock is flat with $\Re_s \rightarrow \infty$, and beyond this, it transitions to a system such that the virtual center lies on the exit plane when observed from SV. Since the opening is asymmetric, the shock depicts ellipsoidal evolution, and the dynamics along the $yz$-plane can be approximated using a model with cylindrical blast geometry with $j=1$.
The blast radius $R_{0}$ for the system outside the tube exit is determined for each case by matching the normalized evolution of blast waves with the model as represented in figure~\ref{fig:shock_outside}d. The corresponding $R_{0}$ values are shown in figure~\ref{fig:shock_outside}e. If we denote $R_0$ values for the system inside and outside the shock tube as $R_{0,in}$ and $R_{0,out}$ respectively, then we can assess the equivalent blast energy $E_{0,in}$ and $E_{0,out}$ using equation~\ref{eq:R_0}. These are presented in figure~\ref{fig:shock_outside}f, where $E_{0,out}$/$E_{0,in}$ represents the fraction of energy of the initial blast energy that is transferred to the blast wave outside, which is approximately $1\%$, irrespective of the blast energy. The rest of the energy is hypothesized to reside with the CVR formation and other secondary shock structures.



\subsection{Induced compressible vortex ring}
\begin{figure}
  \centerline{\includegraphics[width=1\linewidth]{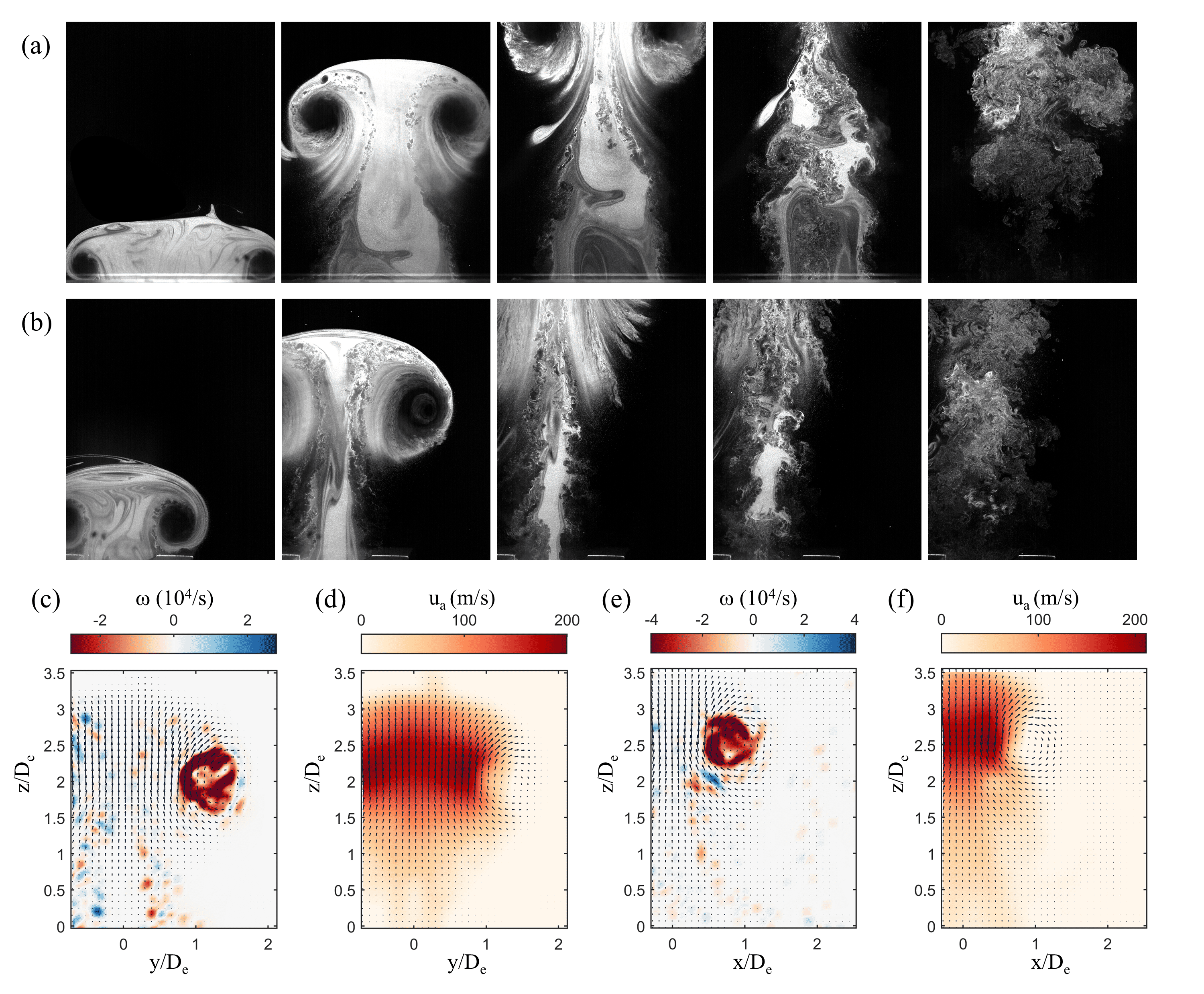}}
  \caption{Compressible vortex ring outside the shock tube at $V_c=6kV$ (a) Mie scattering images of front view (FV) (b) side view (SV). Flow field associated with the flow outside the shock tube, obtained using PIV (c) Vorticity field from FV  (d) axial velocity field from FV (e) Vorticity field from SV  (f) axial velocity field from SV}
\label{fig:vortex_piv}
\end{figure}

The induced flow behind the blast wave is impulsively introduced into the ambient. The flow at the corners conforms to the slipstream, which rolls up to form a CVR. The evolution of the vortex ring is visualized using Schlieren and Mie-scattering imaging as illustrated in figures \ref{fig:schlieren},\ref{fig:phenomenon} and \ref{fig:vortex_piv}.
As a consequence of a rectangular opening, the vortex ring resembles a rectangle in the early stages. The singular kinks arising from the corner are immediately resolved and create waves over the vortex ring as discussed earlier through figure~\ref{fig:phenomenon}.

The flow field associated with these vortex rings is assessed using PIV over two principal planes, namely the $xz$ and $yz$ planes corresponding to the FV and SV, respectively. The obtained flow field is illustrated in Figure \ref{fig:vortex_piv}, showing the vorticity fields $\omega$ and axial velocity $u_a$ ($z$-component). It can be observed that at a given instance along the axis of symmetry, the peak axial velocity is approximately in the middle of the CVR and the velocity decays as we approach the tube opening.

\begin{figure}
  \centerline{\includegraphics[width=1\linewidth]{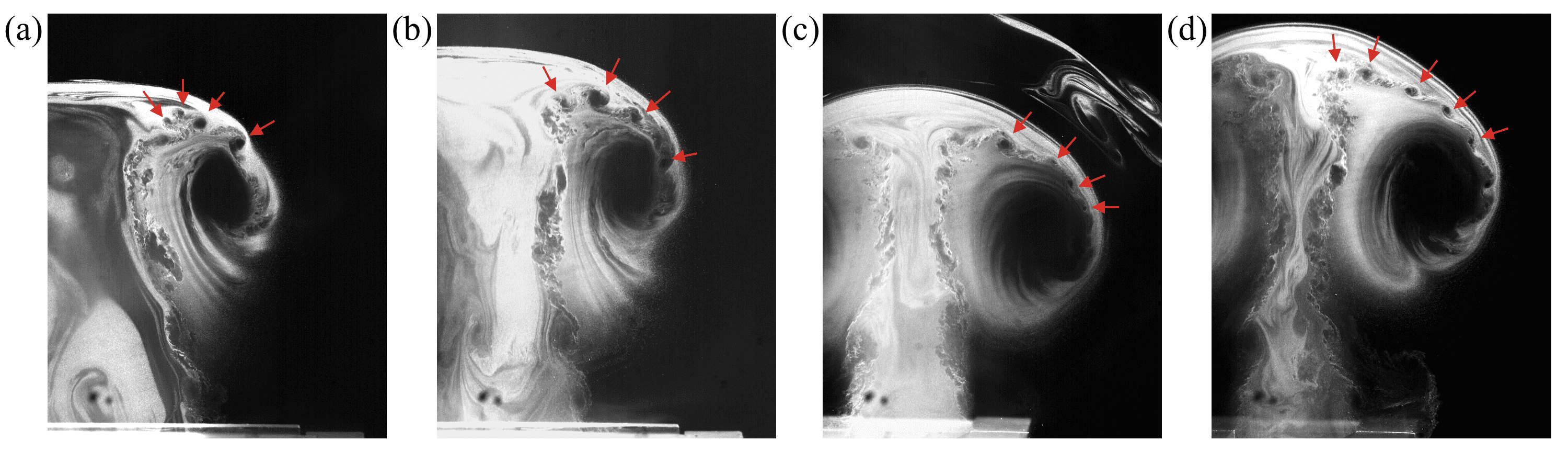}}
  \caption{Mie scattering images of the compressible vortex ring with embedded vortex (red arrows) at higher blast energies (a) $V_c=8kV$ from Front view (FV) (b) $V_c=10kV$ from FV (c) $V_c=8kV$ from Side view (SV) (b) $V_c=10kV$ from SV}
\label{fig:counter_vortex}
\end{figure}

We observe secondary vortex filaments in the schlieren visualization, especially for the flow associated with a high-energy blast in figure~\ref{fig:schlieren}b,d. These can be observed more clearly through planar illumination in the Mie scattering images as illustrated in figure \ref{fig:counter_vortex}, marked with arrows. These secondary vortices are the consequences of the Kelvin-Helmholtz type instability of the slip stream or shear layer that feeds the CVR. The slipstream destabilizes due to unsteady shear and leads to the formation of these convective vortices. As this transient shear layer feeds vorticity to the primary vortex, these secondary vortex structures are transported and engulfed around the primary vortex \citep{doraRoleSlipstreamInstability2014a, thangaduraiCharacteristicsCounterrotatingVortex2010, muruganEvolutionCounterRotating2009}. The shear layer is destabilized beyond certain flow strengths, and hence these vortices appear prominently for higher blast energies.

\begin{figure}
  \centerline{\includegraphics[width=1\linewidth]{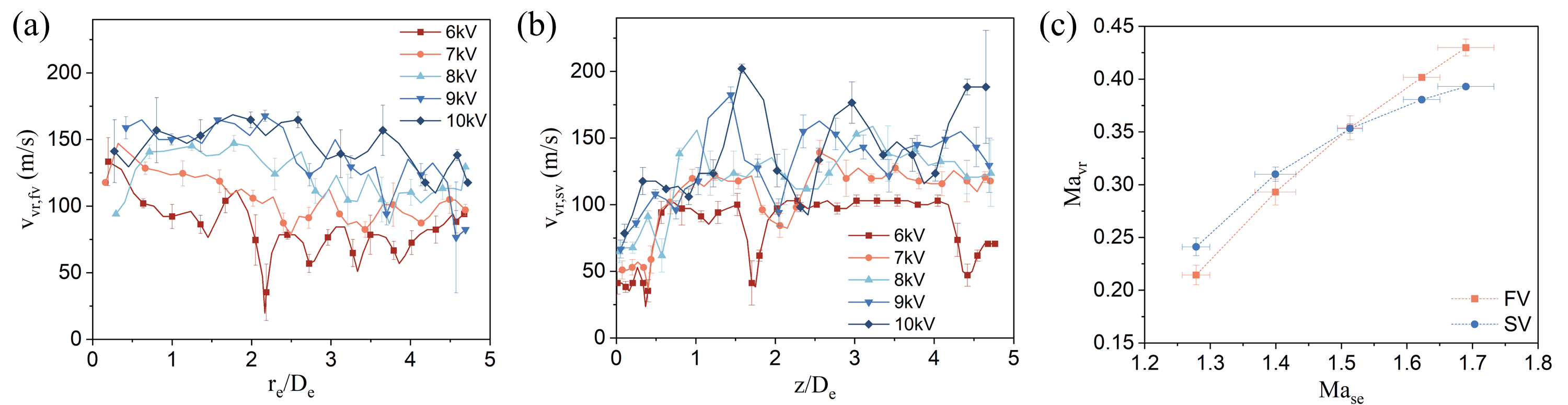}}
  \caption{Temporal evolution of vortex core axial propagation velocity from (a) Front view (FV) (b) Side view (SV) (c) Average convective velocity of vortex core}
\label{fig:vortex_speed}
\end{figure}

The vortex core can be tracked from schlieren images and the $z-t$ diagram shown in figure~\ref{fig:x-t}c,d. This trajectory information is used to determine the temporal evolution of the convective or translational velocity of the cores when viewed from the FV and SV, namely $v_{vr,fv}$ and $v_{vr,sv}$ respectively. We utilized a central difference scheme to estimate the instantaneous velocity, which is shown in figure~\ref{fig:vortex_speed}a,b. A fluctuating motion is observed in CVR translation, as evidenced by figure~\ref{fig:vortex_speed}a; however, more precise measurements are required to enable a definitive analysis. One can observe other localized peaks and dips in the translational velocity in figure~\ref{fig:vortex_speed}a,b, which correspond to the interaction of kinks or waves on the principal planes. This interaction leads to rapid localized movement of the vortex loop due to the superimposition of waves. As apparent from a figure~\ref{fig:x-t}c, when observed over a larger period, CVR exhibits a linear motion. Hence, linear regression was used on the overall trajectory data to determine the average speeds. This was normalized using the speed of sound in the ambient to get $Ma_{vr}=v_{vr}/a_0$, which is illustrated in figure~\ref{fig:vortex_speed}c. The translational velocity of the CVR is observed to increase proportionally with the arrival blast Mach number.

\subsection{Other transient features}

The transient features include a reflected expansion wave that travels inwards and an inward-moving shock wave that follows. There are other secondary shock features observed in the induced flow outside the shock tube. Unlike steady shock cells observed in supersonic jets, in this unsteady transonic jet induced by blast waves, we observe unsteady embedded shocks. These shocklets appear in the central part of the vortex ring, and their recurrent appearance is followed by upstream motion. They disappear, following the reappearance of a subsequent shocklet, and this repeats for a few cycles. This phenomenon is observed for $V_c \geq 7kV$ in the present study, the reason for which will be discussed later. These embedded shocks appear only if the established flow outside the tube is supersonic, and alongside a Prandtl-Meyer fan is established at the opening. These features can be visualized in figure~\ref{fig:schlieren} and are summarized through figure~\ref{fig:x-t}a.


\begin{figure}
  \centerline{\includegraphics[width=0.9\linewidth]{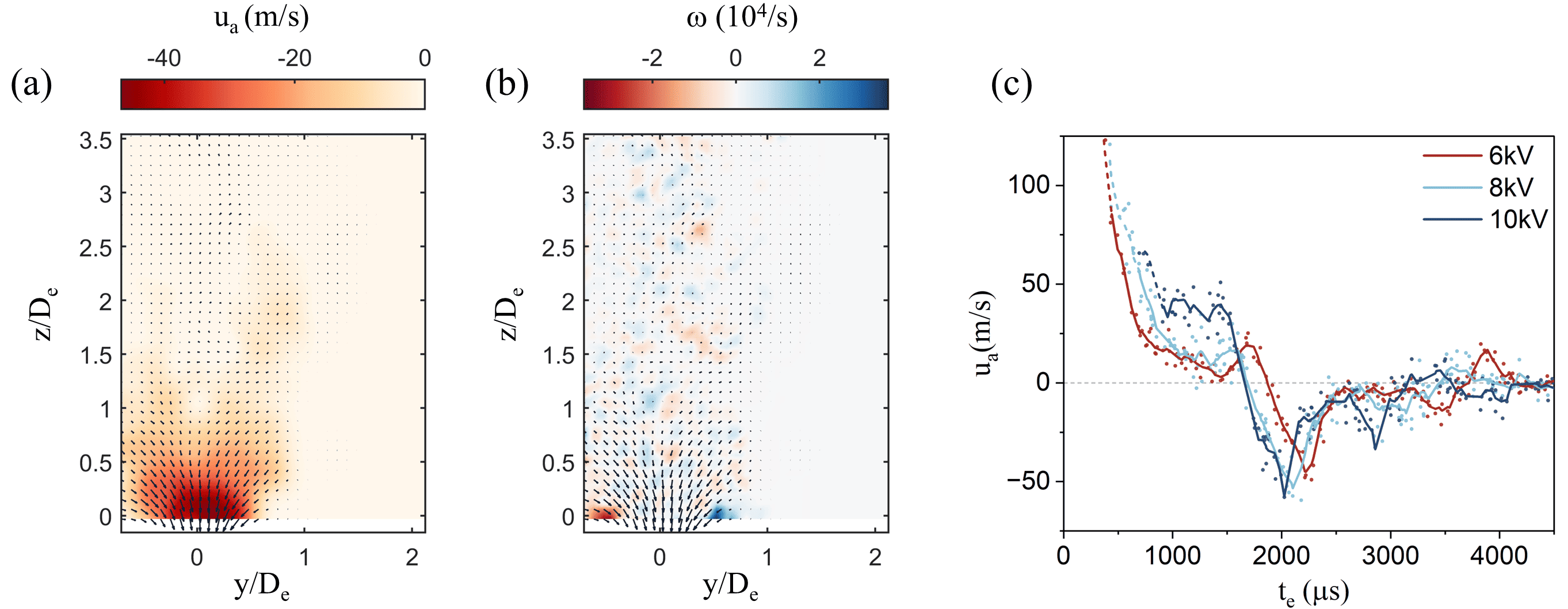}}
  \caption{Flow field of the flow at the shock tube exit illustrating a reversed flow at $V_c = 6kV$ (a) axial velocity field (b) vorticity field (c) Time series of axial velocity at various charging voltages very close to the tube opening at $z/D_e=0.25$}
\label{fig:airflow_reverse}
\end{figure}

The blast wave imposes an unsteady flow that gives rise to these transient features. However, the finite mass of gas inside the shock tube augments the unsteadiness. The shock tube cavity acts as a reservoir of finite size, which is evacuated during the efflux of air. This will eventually be replenished, as evident by the reverse flow observed through PIV measurements, as shown in figure~\ref{fig:airflow_reverse}. This flow reversal is also evident in figure~\ref{fig:x-t}a,b. From figure~\ref{fig:airflow_reverse}a,b, it is observed that the reversed flow is established only in a localized region around the tube opening. At $z=5mm$, i.e., very close to the tube opening, we can trace temporally the axial flow velocity. From the ensembled compilation of measurements from $\sim20$ runs, we deduce the velocity history as illustrated in  figure~\ref{fig:airflow_reverse}c, where the reference time $t_e=t-t_{se}$ with respect to shock arrival is used. The peak reversed values momentarily reach $\sim50m/s$ for the blast energies considered in the present case.

Cumulatively, the blast wave interaction with a shock tube opening encompasses many interesting processes. The mechanisms associated with these prominent transient features will be addressed in the appropriate sections as we proceed to model this phenomenon.


\subsection{Flow field at the shock tube opening}

\begin{figure}
  \centerline{\includegraphics[width=1\linewidth]{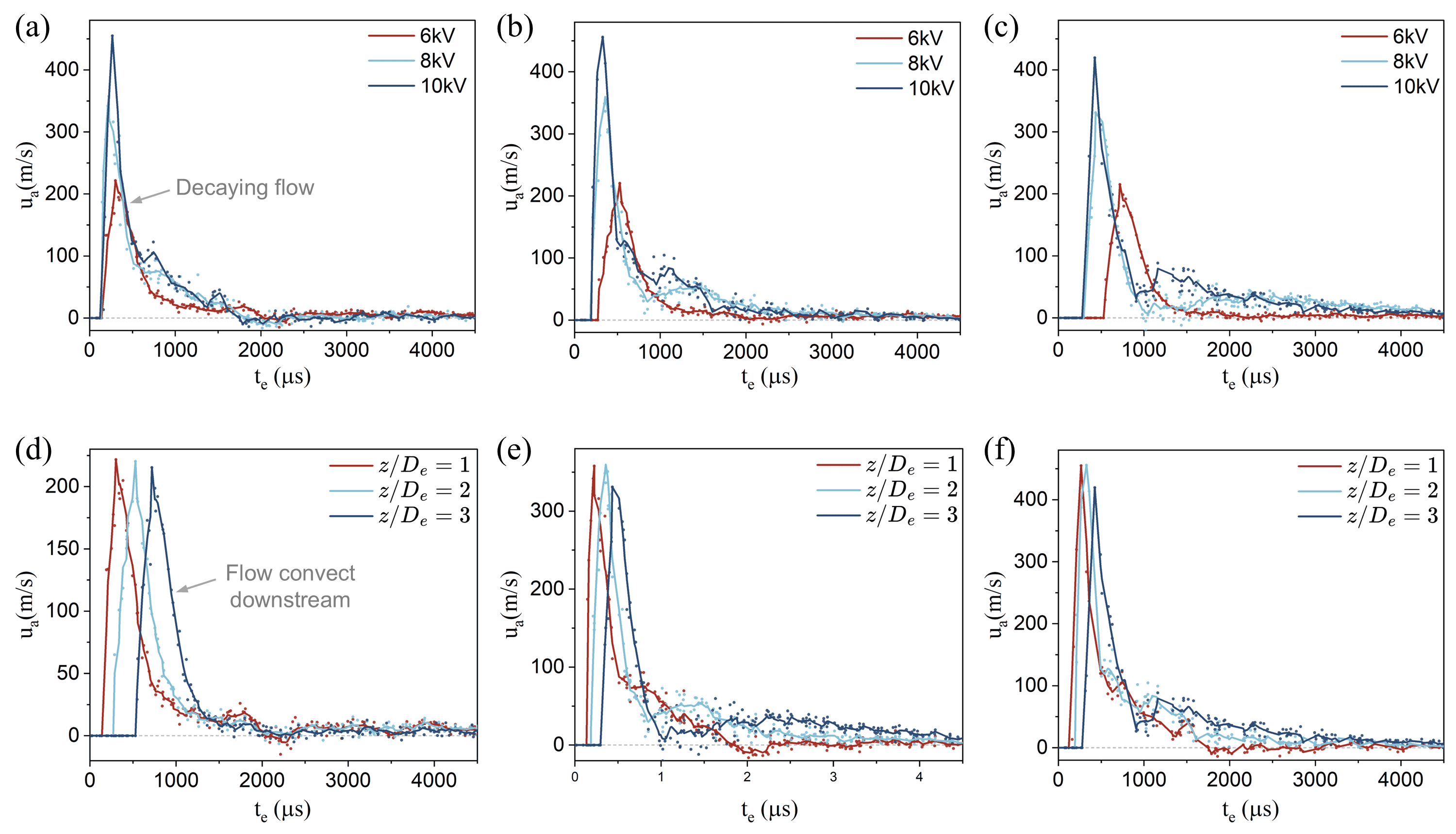}}
  \caption{Evolution of axial flow velocity at a fixed location $z$ from the shock tube exit illustrating a decaying flow field (a) $z/D_e=1$ (b) $z/D_e=2$ (c) $z/D_e=3$. Axial flow velocity for a given blast energy at different $z$ illustrating a convective flow structure (d) $V_c=6kV$ (e) $V_c=8kV$ (f) $V_c=10kV$. These time series data is an ensemble of $\sim20$ runs, measured using PIV.}
\label{fig:airflow_piv}
\end{figure}

We determine the flow field using PIV and consider the flow velocity $u_a$ at various axial locations $z$ along the centerline. The data is not very reliable close to the tube opening at early stages of evolution due to flash from the blast (wire explosion), as discussed earlier. The flow field, as a point measurement on the line of symmetry at $z/D_e=1,2,3$, is obtained from the ensemble of many runs from both FV and SV, and is illustrated in figure~\ref{fig:airflow_piv}. For a given $z/D_e$, figure~\ref{fig:airflow_piv}a-c illustrates a decaying flow profile for different charging voltages $V_c$, where the peak shifts to higher values with the increase in blast energy. The tail of the flow displays a similar decaying trend for all the cases, especially close to the tube opening. Due to the presence of other secondary shock features, the tails start to deviate from each other as we move further downstream. We also observe secondary peaks at higher blast energies, which may be attributed to embedded shock phenomena and reverse flow occurring at the tube opening. For a given $V_c$, figure~\ref{fig:airflow_piv}d-f shows a convective flow profile observed at different $z/D_e$. The peak velocity decays as the CVR propels downstream and loses its strength to various energy loss mechanisms, including viscosity, aero-acoustic interactions, and secondary shock formation.

Using various analytical descriptions, we aim to predict this flow and associated features in this section. The preliminary approach involves the estimation of the flow at the exit plane, neglecting the presence of the tube opening, employing simply the blast wave solutions. The next steps involve assuming a boundary condition associated with the presence of an opening. The various approaches discussed herewith involve a steady outlet pressure (same as ambient pressure) and an acoustics-based impulse response mechanism. The observed flow field and circulation trends are compared with the predictions from these models. In this venture, we deduce a mechanism for other secondary flow features as well.

\subsubsection{Neglecting tube opening effects}

\begin{figure}
  \centerline{\includegraphics[width=1\linewidth]{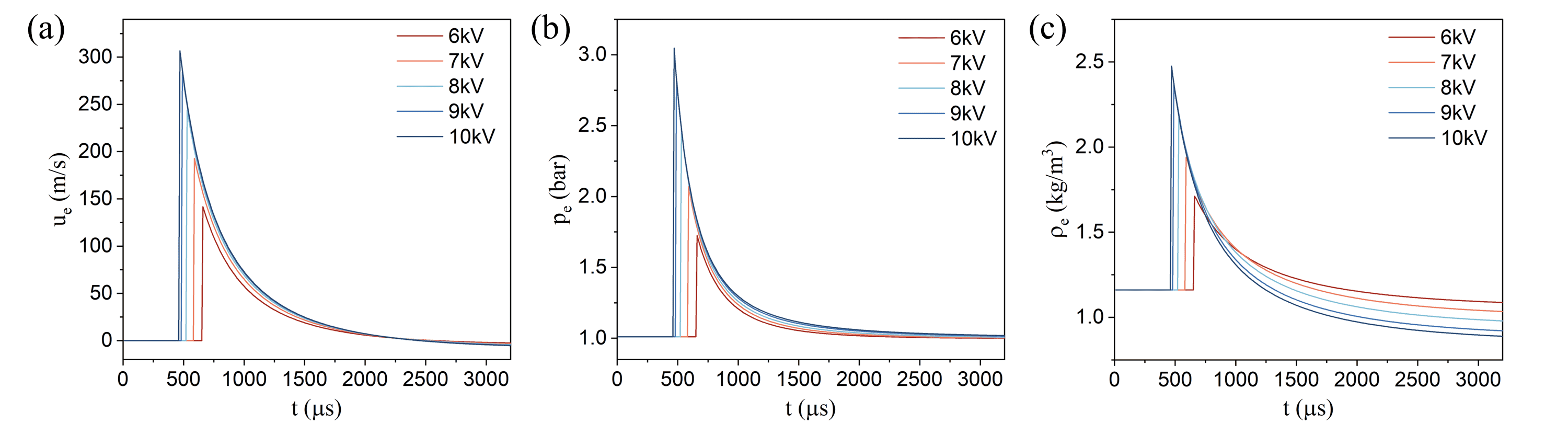}}
  \caption{Evolution of flow parameters at the tube exit location from theoretical blast wave solution ignoring tube opening effects, i.e., assuming an unaffected virtual blast moves downstream  (a) velocity (b) absolute pressure (c) density}
\label{fig:virtual_blast_model}
\end{figure}

Using the model established earlier using the approach presented by \citep{bachAnalyticalSolutionBlast1970}, we can determine the flow at the exit plane of the shock tube. In the first approach, we neglect the opening effects, including artifacts like shock diffraction and inward-moving reflected waves. We solve for the evolution of the unaffected virtual blast wave beyond the tube exit with $R_0$ (or $E_0$) associated with the blast system inside, i.e., $R_0=R_{0,in}$. As discussed earlier, the energy associated with the actual blast system outside the tube is negligible compared to the flow that is produced at the opening. So this virtual blast approach is affordable to a certain extent, if we neglect the tube opening effects.
For the tube exit $\xi=L_{\rm eff}/R_s(t)$ and we solve for the flow parameters $u_e,\ p_e,\ \rho_e$ when $R_s \geq L_{\rm eff}$ or $t \geq t_{se}$, before which $u_e=0,\ p_e=p_0,\ \rho_e=\rho_0$. The obtained flow profiles are presented in figure~\ref{fig:virtual_blast_model}. We observe a step jump corresponding to the appearance of the blast wave and imminent decay of the flow properties, a characteristic associated with such systems. The peak related to the jump elevates in correlation with the increase in blast energy. The tail demonstrates a consistent declining trend across all cases, which aligns with the PIV measurements.

This approach, although obviously erroneous, is simpler to implement and estimate the approximate scales of the flow at the exit plane. In the subsequent sections, we'll use this solution and include the unsteady features accounting for various boundary conditions at the tube opening. Furthermore, for the strong blast wave limit with $Ma_s>>1$, a closed-form solution can be deduced \citep{lee2016gas}, which is extremely convenient to have preliminary estimates and is presented in Appendix~\ref{appA}.

\subsubsection{Effect of tube opening}

\begin{figure}  \centerline{\includegraphics[width=0.9\linewidth]{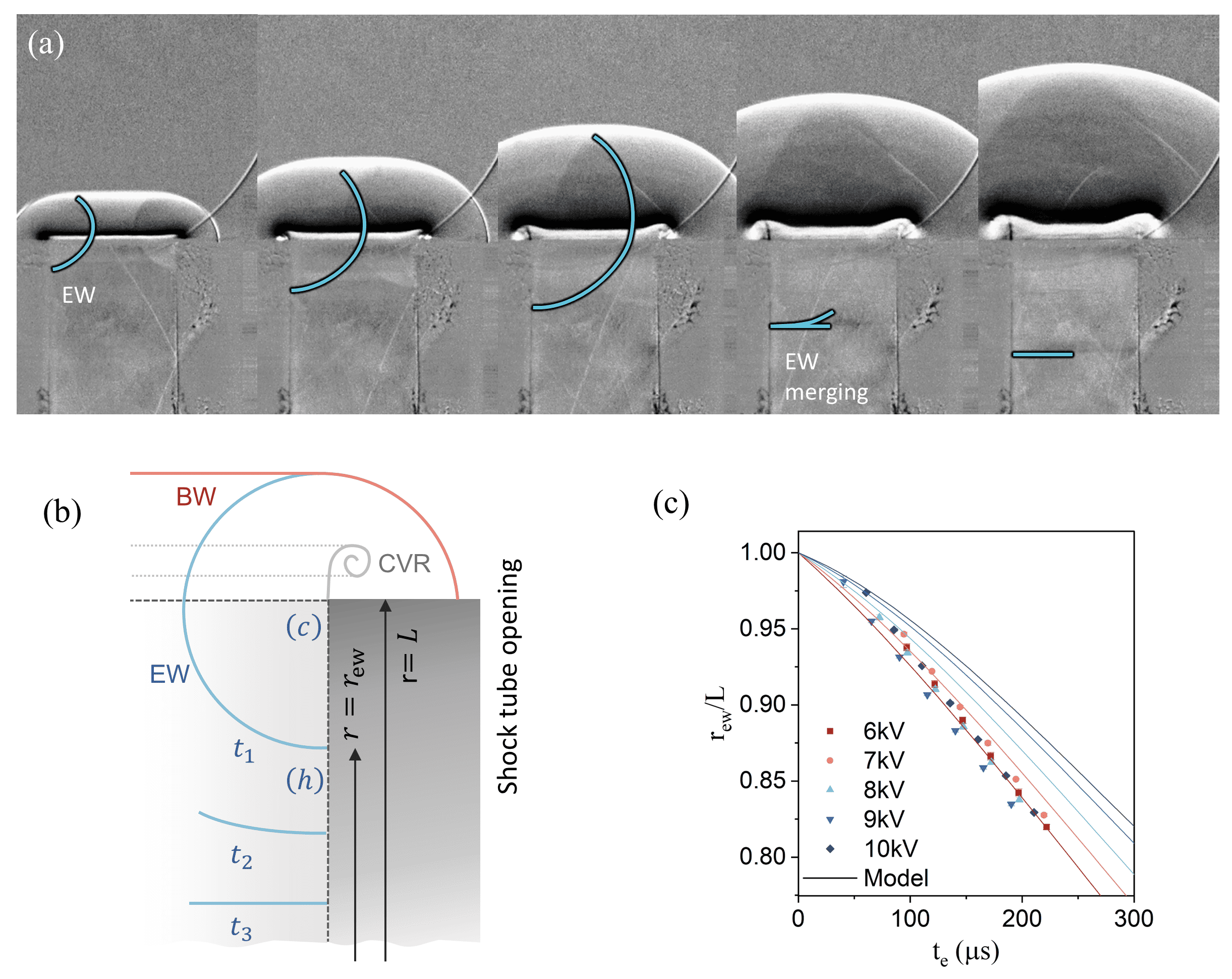}}
  \caption{Schlieren imaging at the tube exit depicting an inwards moving expansion fan (marked with blue line) originating at the corner and merging eventually. (b) Schematic illustrating the diffracting blast wave and expansion wave (c) Trajectory of the expansion wave obtained using Schlieren imaging and comparison with model predictions.}
\label{fig:expansion_wave}
\end{figure}

In the actual system, the presence of an opening or a suddenly expanded area of the flow leads to various flow features, as discussed earlier. The diffraction of the shock wave at the corners or lips generates an inward-moving expansion wave. This expansion fan is centered at the corner, with the head moving inward.  As illustrated in figure~\ref{fig:expansion_wave}a,b, in the early stages, the expansion fan evolves radially with the sharp corner as the center. However, the same phenomenon simultaneously occurs all around the lip periphery. Eventually, all the wave heads merge and morph geometrically into a planar head expansion wave. A similar phenomenon is observed in shock tube flows with partially opened diaphragms \citep{gaetaniShockTubeFlows2008}. Although the early evolution of this system is three-dimensional, it has been shown in the earlier studies that the flow can be treated as one-dimensional near the wall, as inferred from the interferogram \citep{sunFormationSecondaryShock1997, sunVorticityProductionShock2003}. Hence, we can invoke Riemann invariants to determine the flow at the tube exit. We will invoke this approximation throughout the manuscript for the sake of simplicity and to deduce a semi-analytical approach.

\begin{figure}
  \centerline{\includegraphics[width=1\linewidth]{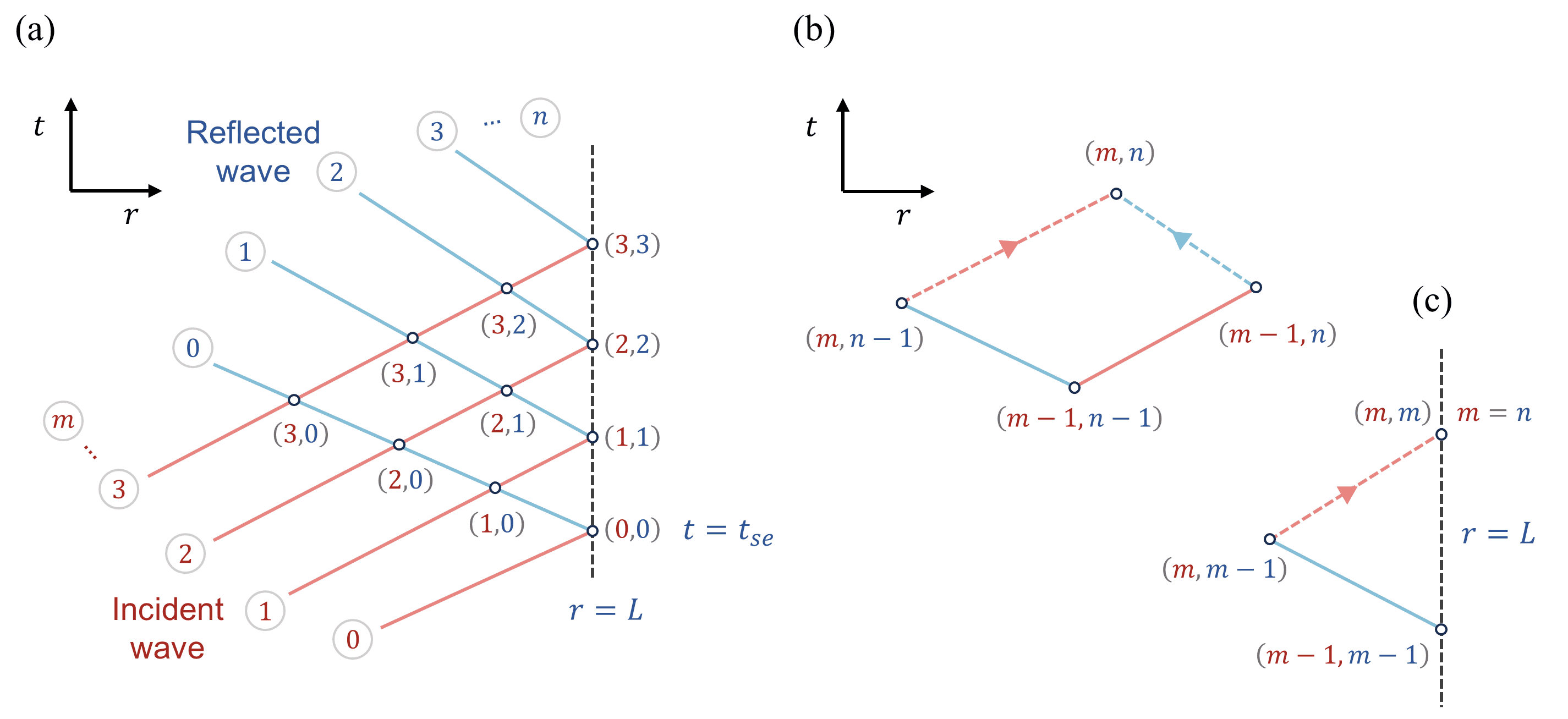}}
  \caption{(a) Schematic illustrating the implementation of the Method of Characteristics, with incident waves (indexed $m$) and reflected wave (indexed $n$) originating from the tube opening at $r=L$, starting at shock arrival time $t=t_{se}$. Characteristics interacting at (b) an intermediate node (c) a wall node with $r=L$.}
\label{fig:moc}
\end{figure}

However, unlike earlier studies involving a simple shock wave, in the present case, the flow upstream at the expansion fan head is unsteady and is denoted by subscript '$h$' as illustrated in figure~\ref{fig:expansion_wave}b. With expansion fan located at $r = r_{\rm ew}$, where $r_{\rm ew}<L_{\rm eff}$ (as it lies inside the tube), we solve for the flow parameters at the wave head $u_h,\ a_h$ at $\xi=r_{ew}/R_s(t)$. However, this wave is a left-running characteristic whose location $r_{ew}$ is determined by integrating

\begin{equation} \label{eq:r_head}
    \frac{d r_{\rm ew}}{dt} = u_h-a_h
\end{equation}

starting from initial conditions $t = t_{se}$ and $r_{\rm ew}=L_{\rm eff}$. The obtained solutions for various blast energies are compared with the experimental trajectories in figure~\ref{fig:expansion_wave}c. The results are in good agreement, indicating the validity of the adapted blast wave solution. To determine the impending flow that follows, incident flow is believed as right-running characteristics that reflect from the tube exit plane following a specified boundary condition, forming left-running characteristics. The reflected wave interacts with the incident waves, which can be tracked in the $r-t$ diagram and can be solved using the method of characteristics (MOC) framework. The Riemann invariants $J$ and slopes $s$ associated with the characteristic curve passing through a point in the $r-t$ diagram depend on the local $u$ and $a$, and are given by \citep{liepmann2001elements, thompson1972compressible}
\begin{eqnarray}
    J^\pm=u\pm\frac{2a}{\gamma-1} \\
    s^\pm=\left(\frac{dr}{dt}\right)^\pm =u\pm a
\end{eqnarray}
where superscripts '$+$' and '$-$' denote the right and left running characteristics, respectively. $J^+$ remains constant throughout the right running characteristics $J^-$ along the left running.
The discretized incident waves can be sequenced with index $m$, starting at $m=0$ corresponding to one just behind the shock wave, and the reflected wave is indexed $n$, as illustrated in figure~\ref{fig:moc}a. The intersection of the incident wave $m$ with the reflected wave $n$ represents a node $\left(m,n\right)$. For these nodes, we can write
\begin{eqnarray}
    u_{m,n}=\frac{1}{2}\left(J_{m,n}^++J_{m,n}^-\right) \\
    a_{m,n}=\frac{\gamma-1}{4}\left(J_{m,n}^+-J_{m,n}^-\right) \\
    J_{m,n}^+=J_{m,n-1}^+=\ldots=J_{m,0}^+ \\
    J_{m,n}^-=J_{m-1,n}^-=\ldots=J_{n,n}^-
\end{eqnarray}
where $J_{m,0}^+$ is determined from the unaltered incident flow and $J_{n,n}^-$ comes from the flow at the tube exit, which requires a boundary condition for closure of this solution. To obtain the position of the node, we utilize the slope of the characteristics, and the previous point in the grid is used to march forward as depicted in figure~\ref{fig:moc}b. As part of the discretized scheme, we assume the characteristics to be constituted of small linear elements between the nodes, and hence we can approximately write $s$ as
\begin{eqnarray}
    s_{m-1,n}^-=\frac{r_{m,n}-r_{m-1,n}}{t_{m,n}-t_{m-1,n}}=u_{m-1,n}-a_{m-1,n} \\
    s_{m,n-1}^+=\frac{r_{m,n}-r_{m,n-1}}{t_{m,n}-t_{m,n-1}}=u_{m,n-1}-a_{m,n-1}
\end{eqnarray}

On simplification, we deduce the coordinates of the next node $(m,n)$ using previous nodes $(m,n-1)$ and $(m-1,n)$ as
\begin{eqnarray}
    t_{m,n}=\frac{\left(r-s^-t\right)_{m-1,n}-\left(r-s^+t\right)_{m,n-1}}{s_{m,n-1}^+-s_{m-1,n}^-} \\
r_{m,n}=r_{m-1,n}+s_{m-1,n}^-\left(t_{m,n}-t_{m-1,n}\right)
\end{eqnarray}
For a node at the boundary, i.e. $r=L_{\rm eff}$ as shown in figure~\ref{fig:moc}c, using a similar argument as above we deduce
\begin{eqnarray}
    s_{m,m-1}^+=\frac{L_{\rm eff}-r_{m,m-1}}{t_{m,m}-t_{m,m-1}}=u_{m,m-1}+a_{m,m-1} \\
    t_{m,m}=t_{m,m-1}+\frac{L-r_{m,m-1}}{s_{m,m-1}^+}
\end{eqnarray}

To determine the flow using this approach, we need to determine the established flow conditions at the outlet to determine the invariants associated with the left-running reflected characteristics. The required boundary condition associated with the tube exit is discussed next.

\subsubsection{Exit boundary conditions – Steady pressure}

In a conventional configuration of flow ejection or a free jet through a nozzle into the ambient, the steady-state solution is determined by imposing an ambient pressure $p_0$ at the opening. However, depending on the initiation of the flow, it is expected that there will be an over-pressure, which is readjusted to the ambient levels over a small but finite amount of time. The compressibility of the flow is expected to enhance these effects. However, for the sake of simplicity, we start with this crude assumption that the exit pressure $p_e=p_0$, which is instantaneously established and is steady throughout the transient outflow.

This assumption, however, is not completely invalid. For the diffraction of a shockwave over a corner, it has been shown by \citet{sunFormationSecondaryShock1997,sunVorticityProductionShock2003} that the pressure at the corner is approximately the same as the ambient pressure, especially for the strong shocks. This led them to establish flow conditions at the corner and the extended neighborhood using this assumption and isentropic flow conditions. For simplicity, we extend their approach and incorporate it into the MOC implementation. 

For a node $\left(m,m\right)$ at the boundary, conditions at node $\left(m,m-1\right)$ are used to determine the flow conditions (see figure~\ref{fig:moc}c) using the Riemann invariants, we get $J_{m,m}^+=J_{m,m-1}^+$ or
\begin{equation} \label{eq:riemann_p0}
    u_{m,m}+\frac{2a_{m,m}}{\gamma-1}=u_{m,m-1}+\frac{2a_{m,m-1}}{\gamma-1}
\end{equation}

By imposing isentropic relations  \citep{liepmann2001elements, thompson1972compressible} between these nodes, we get
\begin{equation} \label{eq:isentropic_p0}
    \frac{a_{m,m}}{a_{m,m-1}}=\left(\frac{p_{m,m}}{p_{m,m-1}}\right)^\frac{\gamma-1}{2\gamma}
\end{equation}

Substituting this in equation~\ref{eq:riemann_p0} and simplifying
\begin{equation} \label{eq:outflow_p0}
    u_{m,m}=u_{m,m-1}+\frac{2a_{m,m-1}}{\gamma-1}\left\{1-\left(\frac{p_{m,m}}{p_{m,m-1}}\right)^\frac{\gamma-1}{2\gamma}\right\}
\end{equation}

Finally, substituting the steady outlet pressure condition 
\begin{equation}
    p_{m,m}=p_0
\end{equation}

$u_{m,m}$ and $a_{m,m}$ can be determined from the equations~\ref{eq:isentropic_p0} and \ref{eq:outflow_p0}. This prescribes the exit flow Mach number ${Ma}_e$ as 
\begin{equation}
    {Ma}_e\equiv{Ma}_{m,m}=\frac{u_{m,m}}{a_{m,m}}
\end{equation}

\begin{figure}
  \centerline{\includegraphics[width=0.7\linewidth]{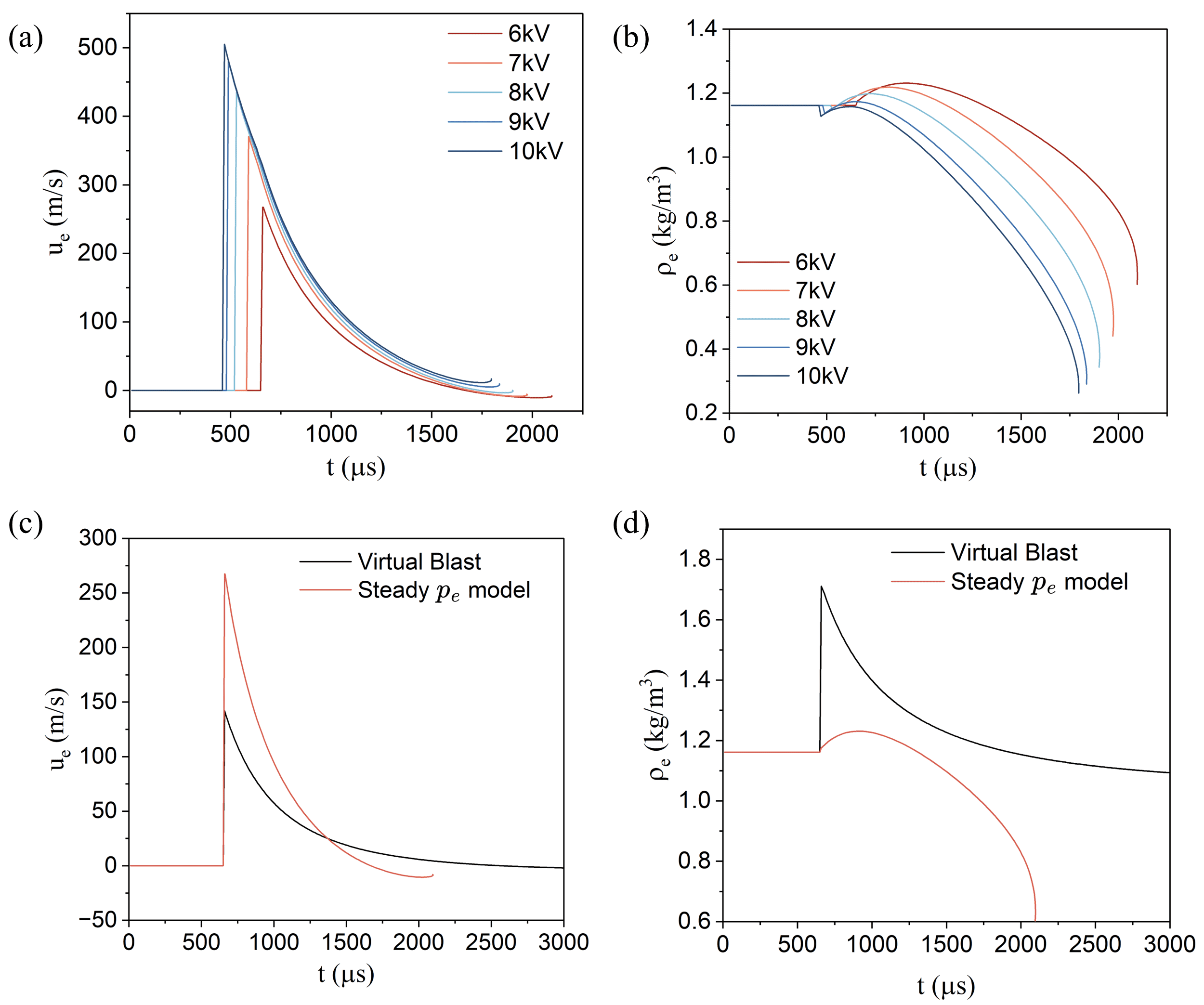}}
  \caption{Evolution of flow parameters at the tube exit considering the opening effects with a steady pressure model (a) velocity (b) density. Comparison of this model with the virtual blast model (c) velocity (d) density}
\label{fig:steady_p_model}
\end{figure}

If ${Ma}_{m,m}\geq1$, the validity of the left-running characteristics within the tube is compromised. Accompanying this supersonic flow, a Prandtl-Meyer (PM) expansion fan appears at the exit of the shock tube, centred at the corner. We assume the sonic characteristic of the PM fan to be aligned with the tube exit plane, i.e., perpendicular to the wall \citep{thompson1972compressible,sunVorticityProductionShock2003}. Corresponding to this configuration, the Mach number at the corner before the PM fan should be ${Ma}_c=1$, and in the previous equations, we assume
\begin{equation}
    {Ma}_{m,m}\equiv{Ma}_c=1
\end{equation}
Henceforth, subscript '$c$' corresponds to the flow at the corner and '$e$' for the flow after the PM fan. Substituting this in equation~\ref{eq:riemann_p0}, we get
\begin{equation}
    u_{m,m}=a_{m,m}=\frac{\gamma-1}{\gamma+1}\left\{u_{m,m-1}+\frac{2a_{m,m-1}}{\gamma-1}\right\}
\end{equation}
and $p_c\equiv\ p_{m,m}$ can be deduced from equation~\ref{eq:isentropic_p0}. To determine the outflow conditions, we invoke our steady pressure assumption $p_e=p_0$ and isentropic relations across the PM fan as follows  \citep{liepmann2001elements, thompson1972compressible}
\begin{equation}
    \frac{p_e}{p_c}=\left\{\frac{1+\frac{\gamma-1}{2}{Ma}_c^2}{1+\frac{\gamma-1}{2}{Ma}_e^2}\right\}^\frac{\gamma}{\gamma-1}
\end{equation}
\begin{equation} \label{eq:isentropic_p0}
    \frac{a_{e}}{a_{c}}=\left(\frac{p_{e}}{p_{c}}\right)^\frac{\gamma-1}{2\gamma}
\end{equation}
We can solve for the outflow Mach number ${Ma}_e$ as
\begin{equation}
    {Ma}_e^2=\frac{\gamma-1}{\gamma+1}\left(\frac{p_e}{p_c}\right)^\frac{\gamma-1}{2}-\frac{2}{\gamma-1}
\end{equation}
and subsequently velocity ${u}_e = {Ma}_e \cdot c_e$. For $V_c=6kV$, the ${Ma}_{m,m}<1$ throughout, and hence the solution doesn't involve any PM fan, as evident from the experiments as well. For $V_c\geq7kV$, we get ${Ma}_{m,m}\geq1$ for a short duration in which we solve for the PM fan as discussed earlier and do not consider any inward-moving characteristics. The instance when the exit flow becomes subsonic, we evaluate the following flow using MOC with left-running characteristics inside the tube. The corrected flow at the tube exit, considering a reflected expansion wave, is then $u_{e} \equiv u_{m,m},\ p_{e} \equiv p_{m,m} = p_0,\ \rho_{e} \equiv \rho_{m,m}$ at times $t \equiv t_{m,m}$; as presented in figure~\ref{fig:steady_p_model}a,b. This solution is compared with the virtual blast wave solutions obtained earlier in figure~\ref{fig:steady_p_model}c,d. The peak velocity post impulsive jump has significantly increased, and this is expected since instantaneously normalizing the overpressure to zero must be accompanied by elevated velocities by virtue of conservation principles. The density is found to be lower, and this is expected by virtue of mass conservation.

\begin{figure}
  \centerline{\includegraphics[width=1\linewidth]{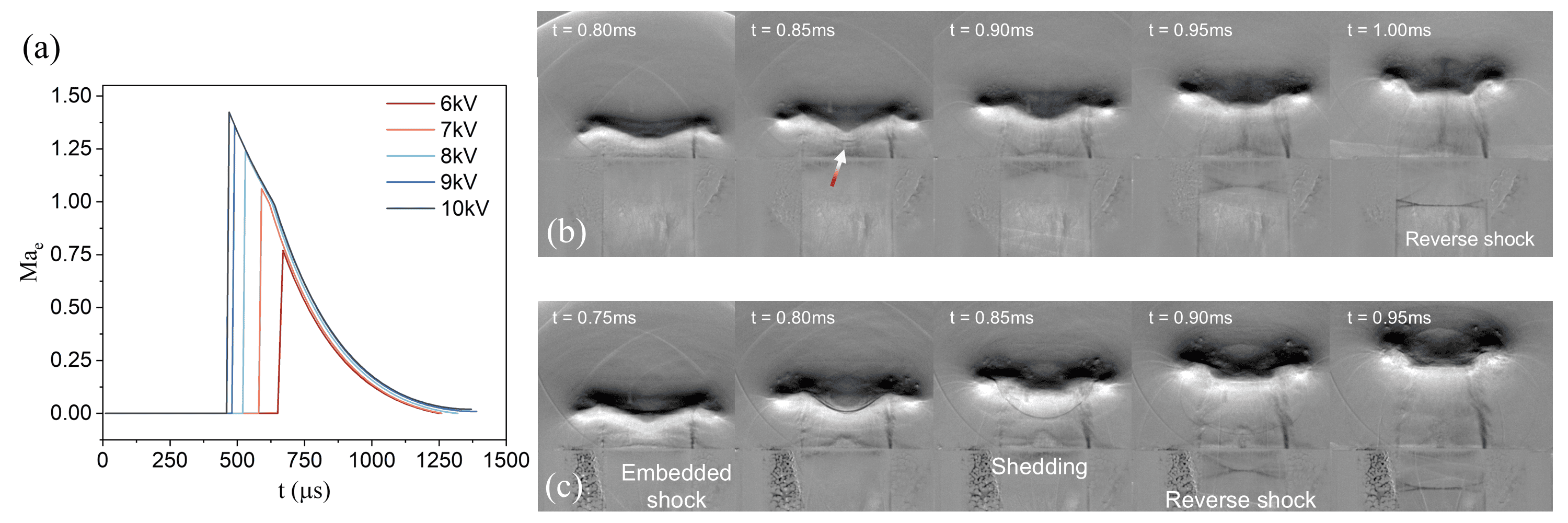}}
  \caption{(a) Evolution of flow Mach number at the tube exit for the model considering the opening effects with a steady pressure model. (b) Temporal evolution of embedded shock structures in the compressible vortex ring, illustrating their appearance beyond a threshold blast energy (b) $V_c=7kV$ (c) $V_c=8kV$}
\label{fig:steady_p_Ma}
\end{figure}

The evolution of the flow Mach number $Ma_{e}$ at the tube exit is presented in figure~\ref{fig:steady_p_Ma}a. It can be observed that for $V_c=6kV$, the flow is always subsonic, i.e., $Ma_{e}<1$. However, for higher voltages, we observe a supersonic flow for a short time duration, and it is expected that shock structures should appear in the flow downstream. This hypothesis aligns well with the experimental observations, where we see embedded shock structures in the compressible vortex ring and its trailing jet only for $V_c\geq7kV$, as illustrated in figure~\ref{fig:steady_p_Ma}. We also see that the shock appears for a very short duration for $V_c=7kV$ (marked with a red arrow in figure~\ref{fig:steady_p_Ma}b) as is expected from the obtained solutions, where $Ma_{e}>1$ for a very small period.
The appearance of such structures beyond a certain threshold aligns well with the theoretical solutions. A similar threshold has been predicted earlier by \cite{sunFormationSecondaryShock1997} for the appearance of secondary shocks during the shockwave diffraction over a corner for an incident shock with shock Mach number ${Ma}_s>1.34$. This condition is fulfilled in the current experiments as well.

\begin{figure}
  \centerline{\includegraphics[width=1\linewidth]{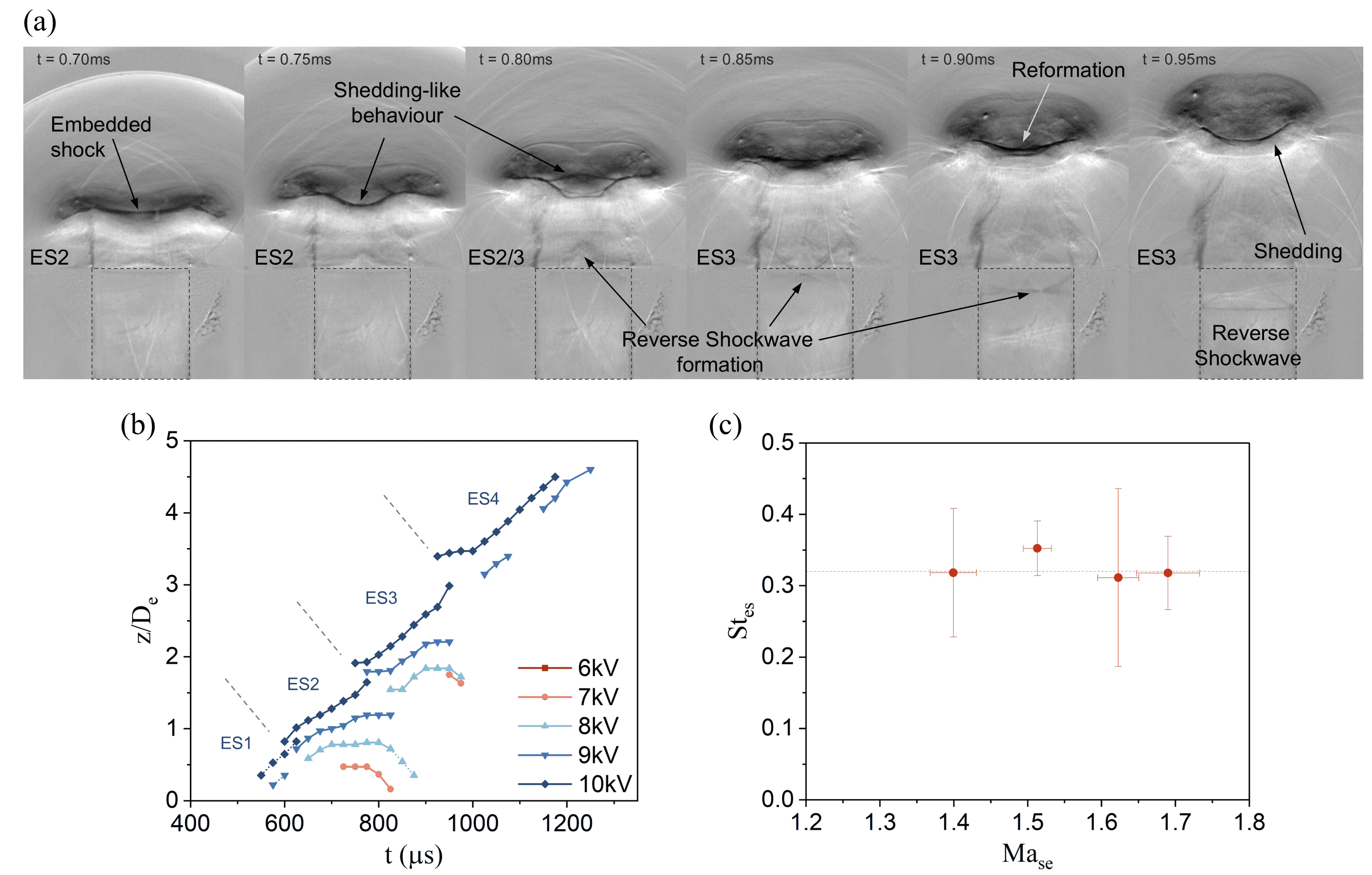}}
  \caption{(a) Evolution of embedded shock (ES) structures in the compressible vortex ring, illustrating a shedding pattern at $V_c=10kV$ (b) Trajectories of the ES on the $z-t$ diagram, with 'ES$i$' indicating the $i$-th generation (c) Strouhal number $(St_{es})$ associated with the ES wave shedding. }
\label{fig:embedded_shock}
\end{figure}

These embedded shocks appear and disappear repeatedly, demonstrating a shedding-like behaviour as shown in figure~\ref{fig:embedded_shock}a. These embedded shocks are tracked in the $z-t$ space as illustrated in figure~\ref{fig:embedded_shock}b, and ES$i$ represents the $i$-th shock generation. The average time period of the shedding $T_{\rm es}$ is measured from this diagram, and the associated frequency is defined as $f_{\rm es} = 1/T_{\rm es}$.
The rapid decay of outflow, along with the unsteady nature of vortex formation, might be the reason for this unsteady shedding behavior. Another possible mechanism entertains the propagation of the disturbances from the tube opening along the shear layer that feeds vorticity to the vortex ring in the early stages \citep{ thasuStrouhalNumberUniversality2022,thasuAeroacousticMechanismsExplain2025,songRoleConvectingDisturbances2025}. These disturbances might act as an unsteady perturbation that persists after the vortex pinch-off as well. Analogous recurrent phenomena are observed within a transonic diffuser, characterized by the development of multiple localized shock waves  or shocklets \citep{handaFormationMultipleShocklets2002}. The disturbance scale is expected to be represented by the tube cross-section dimension $D_e$ and the maximum speed of sound at the outlet ${\rm max}(a_{e})$ (just after shock wave incidence). Using these scales to normalize the obtained shedding frequency, we can obtain a shedding Strouhal number ${St}_{\rm es}$, defined as 
\begin{equation}
    {St}_{\rm es} = \frac{f_{\rm es}D_e}{{\rm max}(a_{e})} 
\end{equation}

The shedding Strouhal number is found to be universal for all the cases in the present study, ${St}_{\rm es} \approx 0.32$, as illustrated in figure~\ref{fig:embedded_shock}c. This is evident in figure~\ref{fig:embedded_shock}b as well, where we see a pattern common to all the cases. Further investigation is necessary to establish the exact mechanism for this instability and universality.

\begin{figure}
  \centerline{\includegraphics[width=0.8\linewidth]{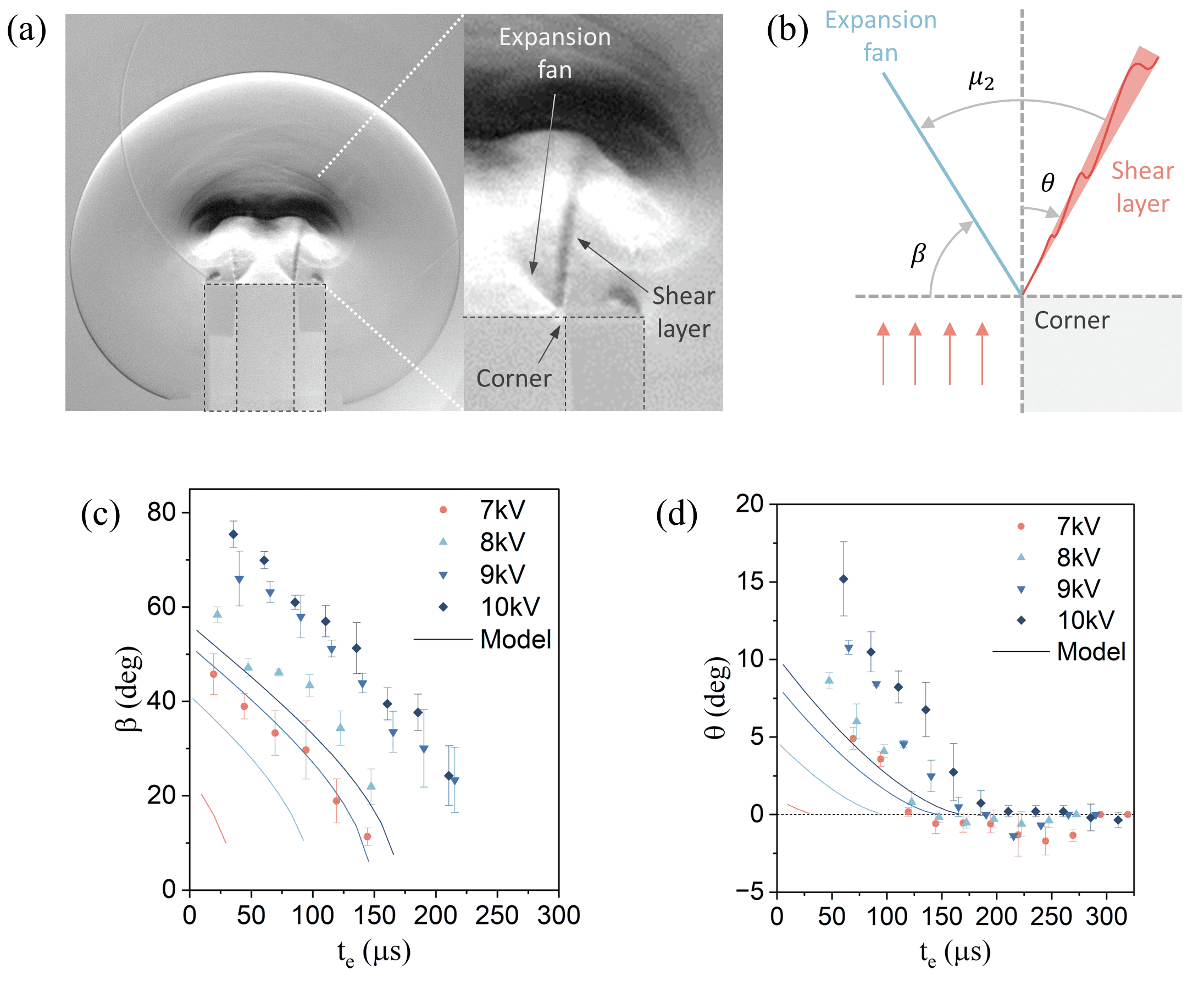}}
  \caption{(a) A close-up of the flow field generated at the shock tube opening from side view, displaying a shear layer and Prandtl-Meyer (PM) oblique expansion fan near the corner (b) Schematic of the PM wave system illustrating angular positions of prominent structures. Angular position of (c) PM expansion fan $\beta$ and (d) shear layer $\theta$ outside the tube and comparison with the steady pressure model predictions. }
\label{fig:steady_p_PM}
\end{figure}

The presented flow model can also account for the transient shear layer and right-most characteristic of the PM fan visible outside the shock tube in the experiments, as shown through SV in figure~\ref {fig:steady_p_PM}a. Using the standard PM functions, we can interpret the geometrical properties of the flow, i.e., deflection angle $\theta$ as  \citep{liepmann2001elements, thompson1972compressible}
\begin{equation}
    \theta=\nu\left({Ma}_e\right)-\nu\left({Ma}_c\right)
\end{equation}

where $\nu$ is the PM function. The deflection angle physically corresponds to the location of the shear layer as illustrated in figure ~\ref{fig:steady_p_PM}b. The flow doesn't follow the actual tube geometry due to the pressure boundary condition and separates at the corner, forming a shear layer (ideally an infinitesimally thin slipstream theoretically). The observed position of the shear layer, obtained from schlieren images, is compared with the theoretical prediction in figure ~\ref{fig:steady_p_PM}c. The position of the last PM wave can be determined from this model as
\begin{equation}
    \beta = 90^o+\theta-\mu_e
\end{equation}
where $\mu_e = \sin^{-1}\left( 1/{Ma}_e\right)$ is the downstream Mach angle. The  PM wave edge observed in experiments is compared with the model in figure ~\ref{fig:steady_p_PM}d. The predictions are in good agreement with experiments, considering the simplicity of the model with the imposed steady ambient pressure condition.

\begin{figure}
  \centerline{\includegraphics[width=0.9\linewidth]{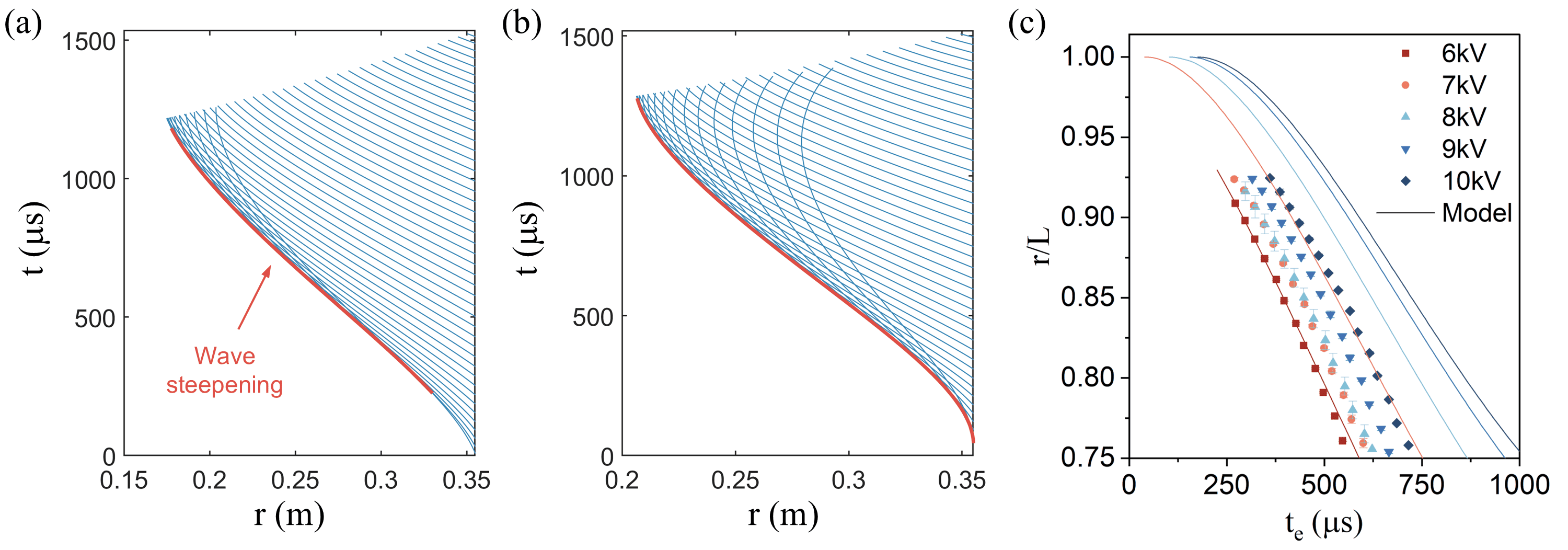}}
  \caption{Left running characteristics (blue) obtained from the steady pressure model depicting steepening and formation of an inward moving shock marked with red lines (a) $V_c=6kV$ (b) $V_c=7kV$. (c) Comparison of the shock trajectories from the steady pressure model with experiments.}
\label{fig:steady_p_reshock}
\end{figure}

Another peculiar observation involves the formation of an inward-moving shock wave or reverse shock or "reshock" originating near the tube opening during the process of reflection of the blast wave-induced flow. This reshock is observed in all the cases, and the trajectory is obtained from schlieren images. To assess the presence of such discontinuities in the proposed model, we look closely at the geometrical nature of the obtained characteristics. Figure~\ref{fig:steady_p_reshock}a-b shows the left-running reflected characteristics obtained through MOC. For $V_c=6kV$, we see that beyond a certain point, the flow steepens and characteristics come together and eventually cross each other. This steepening forms an envelope marked with a red line. The region with crossing characteristics is not physically possible because this region then corresponds to a multivalued solution. For a well-behaved solution, a discontinuity must be introduced, and it is expected that this envelope represents a shock wave. For $V_c\geq7kV$, by virtue of a transonic flow at the tube opening, this shock envelope originates at the exit and propagates inwards as illustrated in figure~\ref{fig:steady_p_reshock}b. Shock trajectories deduced from this model are plotted with the experimental data in figure~\ref{fig:steady_p_reshock}c, and a good agreement is realised. Although the exact mechanism of introducing the shock and interaction of the characteristics with it is not considered in the present study, it will be taken up in the future. However, the general approximations introduce a discontinuity within the region where characteristics cross without affecting the rest of the flow \citep{whitham2011linear}. Extending this premise, we expect the flow estimated at the opening to be consistent with or without reshock. Additionally, on a similar note, reflection of an expansion wave from an open end of the tube leads to reflected compression waves with convergence or steepening leading to a reverse shock wave \citep{kimPropagationCharacteristicsCompression2003}. The blast wave is synonymous with a shock wave followed by a diverging wave or expansion fan \citep{lee2016gas}, and hence the following characteristics, when reflected, steepen in a similar fashion to form the reshock.

Although this boundary condition of $p_e=p_0$ is extremely simple to implement, there are several obvious drawbacks associated with the assumptions. This will be discussed later in the upcoming sections.

\subsubsection{Exit boundary conditions – Impulse response}

The earlier solution employed a major assumption that the effective pressure is the same as the ambient pressure associated with the steady flow conditions.
In actual systems, however,  it is expected that the steady flow conditions are asymptotically approached after a lag period depending on the strength of the incident disturbance or, in this case, a blast wave and the accompanying unsteady flow. \cite{rudingerReflectionPressureWaves1957, rudingerNonsteadyDischargeSubcritical1961} used an acoustic theory based on impulse response to deduce the flow conditions associated with the reflection of shock waves and arbitrary pressure waves for the open end of a tube. This results in an overpressure at the tube exit in response to the shock wave (or pressure wave), which subsequently dissipates quickly to ambient pressure within a time frame comparable to the duration required for a sound wave to propagate through a few duct diameters. It was shown that the instantaneous flow conditions, illustrated in terms of the speed of sound, depend on the integrated history. Using this approach for an arbitrary pressure wave \citep{rudingerReflectionPressureWaves1957}, we have
\begin{equation} \label{eq:acoustic_int}
    a_n=a_0+\sum_{j=1}^{n}{\frac{\Delta a_j}{\tau_j-\tau_{j-1}}\left\{\Phi\left(\tau_n-\tau_{j-1}\right)-\Phi\left(\tau_n-\tau_j\right)\right\}}
\end{equation}

where
\begin{eqnarray}
    a_i\equiv\ a_{i,i},\ \ \tau_i\equiv\tau_{i,i}=\frac{t_{i,i}a_0}{D_e} \\
    \Delta a_i=a_{i,0}-a_{i-1,0}
\end{eqnarray}

We substitute $\Delta a (\tau)$ from the incident blast wave solutions. The function $\Phi$ was defined as 
\begin{equation}
    \Phi\left(\tau\right)=\int_{0}^{\tau}{I\left(\tau^\prime\right)d\tau^\prime}
\end{equation}

Where $I(\tau)$ is an impulse response function
\begin{equation}
    I\left(\tau\right)=\frac{a_{act}\left(\tau\right)-a_0}{a_{inc}-a_0}
\end{equation}

With $\left(a_{inc}-a_0\right)$ denoting a jump in speed of sound across the incident wave, and $(a_{act}\left(\tau\right)-a_0)$ is the actual jump observed. The functions $\Phi$ and $I$ are obtained from tables provided by \cite{rudingerReflectionPressureWaves1957}, which were obtained from acoustic principles. Hence, as evident from equation~\ref{eq:acoustic_int}, this approach involves a complete history of the incident flow to determine the established flow at the outlet. To determine other flow parameters, the initial stationary condition of flow outside the tube is used, and using Riemann invariants and isentropic conditions, \citep{rudingerReflectionPressureWaves1957}
\begin{eqnarray}
    \frac{p_e}{p_0}=\left(\frac{a_e}{a_0}\right)^\frac{2\gamma}{\gamma-1} \\
    u_e=\frac{2}{\gamma-1}\left(a_e-a_0\right)
\end{eqnarray}

\begin{figure}
  \centerline{\includegraphics[width=1\linewidth]{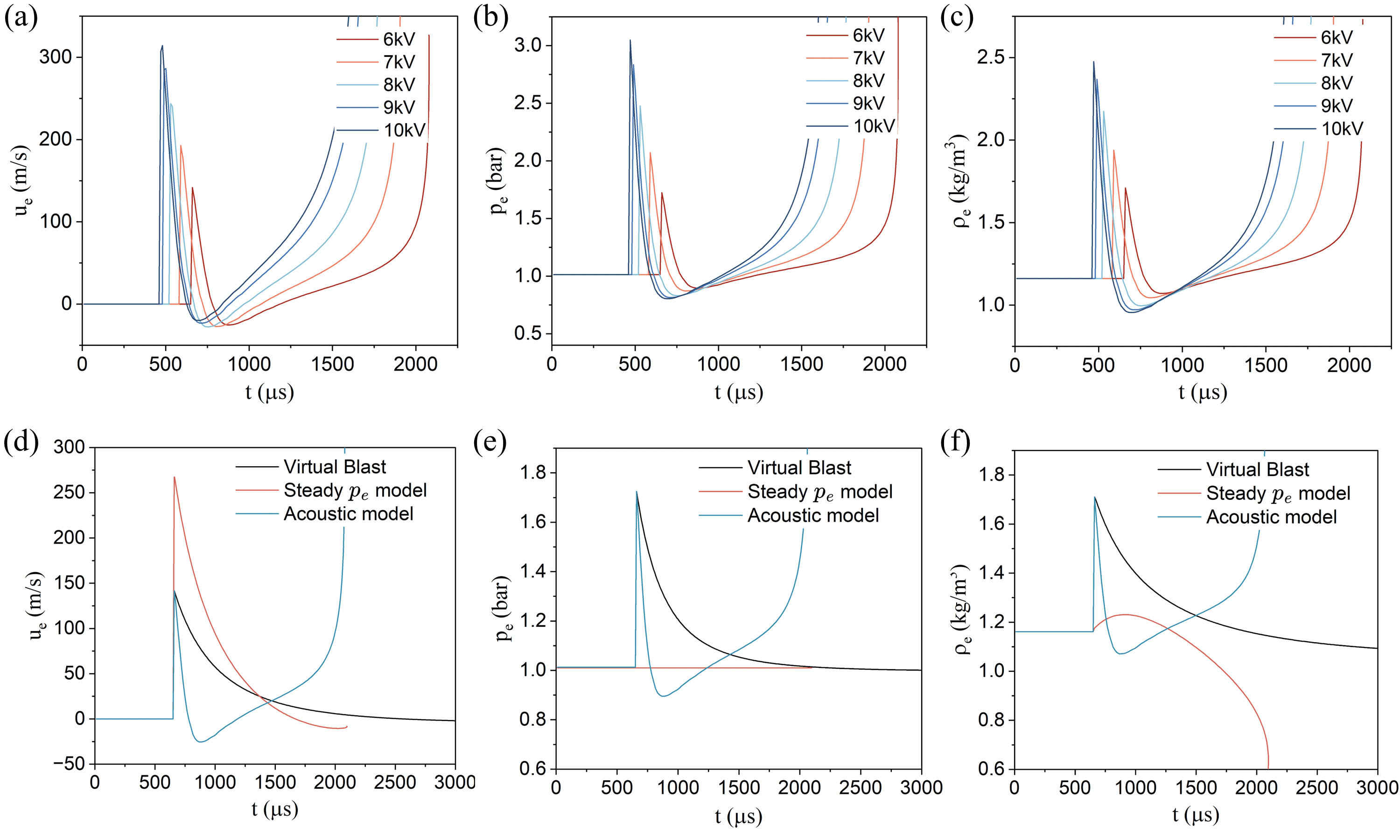}}
  \caption{Evolution of flow parameters at the tube exit considering the opening effects with an acoustic model (a) velocity (b) pressure (c) density. Comparison of this model with the virtual blast model (d) velocity (e) pressure (f) density.}
\label{fig:acoustic_model}
\end{figure}

\begin{figure}
  \centerline{\includegraphics[width=0.7\linewidth]{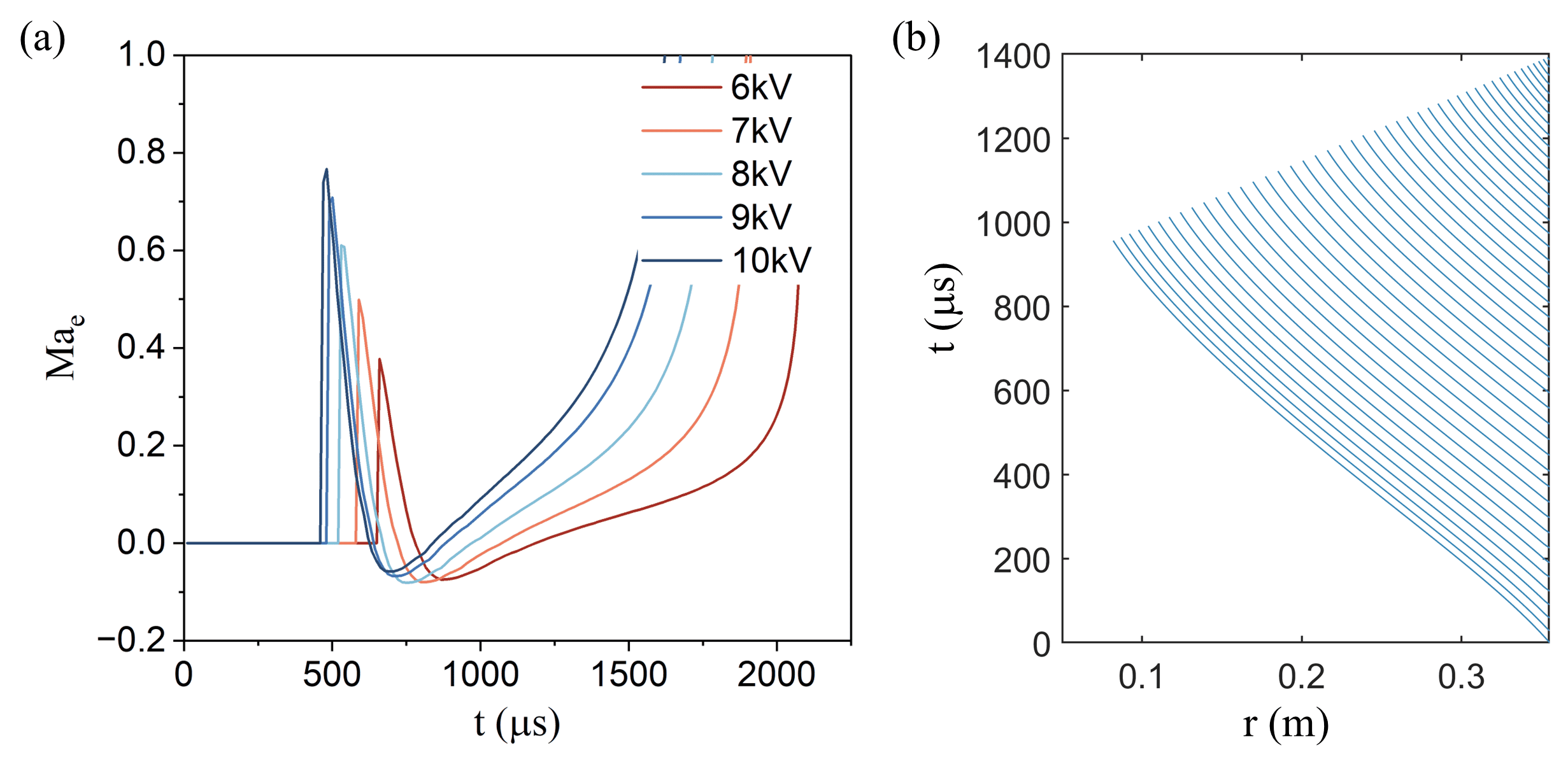}}
  \caption{(a) Evolution of flow Mach number at the tube exit for the model considering the opening effects with an acoustic model. (b) Left running characteristics (blue) derived from the acoustic model lacking steepening or an inward-moving shock ($V_c=6kV$). }
\label{fig:acoustic_Ma}
\end{figure}

The values of $a_e$ and $u_e$ are incorporated in the earlier discussed MOC approach at tube boundary nodes to obtain the complete wave diagram. The obtained solutions are illustrated in figure~\ref{fig:acoustic_model}a-c. We observed a peak after the shock passes, flow decays, and eventually a flow reversal is also predicted with negative $u_e$ values as well. The obtained flow reversal velocities are in close agreement with the values obtained through PIV measurement, as illustrated earlier in figure~\ref{fig:airflow_reverse}b. The solutions eventually blow up because the rarefaction head reaches the other end of the shock tube, and the velocity of sound (and temperature) blows up for the adapted blast wave solution at $r\rightarrow0$. The comparison of solutions obtained with other boundary conditions is illustrated in figure~\ref{fig:acoustic_model}d-f. Compared to the solutions neglecting the tube boundary altogether, the peak matches, as expected \citep{rudingerReflectionPressureWaves1957}, and the trailing flow decays rather quickly. This decay is even quicker than the experimental observations. This approach predicts subsonic exit flow Mach number ${Ma}_e<1$ for all the blast energies as shown in figure~\ref{fig:acoustic_Ma}a, which is inconsistent with the appearance of embedded shock and necessity of supersonic out flow. Also, the left running characteristics do not cross, as illustrated in figure~\ref{fig:acoustic_Ma}b. Without steepening of the wave, this theory fails to predict the presence of reshock.

\subsubsection{Model validation and vortex dynamics}

\begin{figure}
  \centerline{\includegraphics[width=1.0\linewidth]{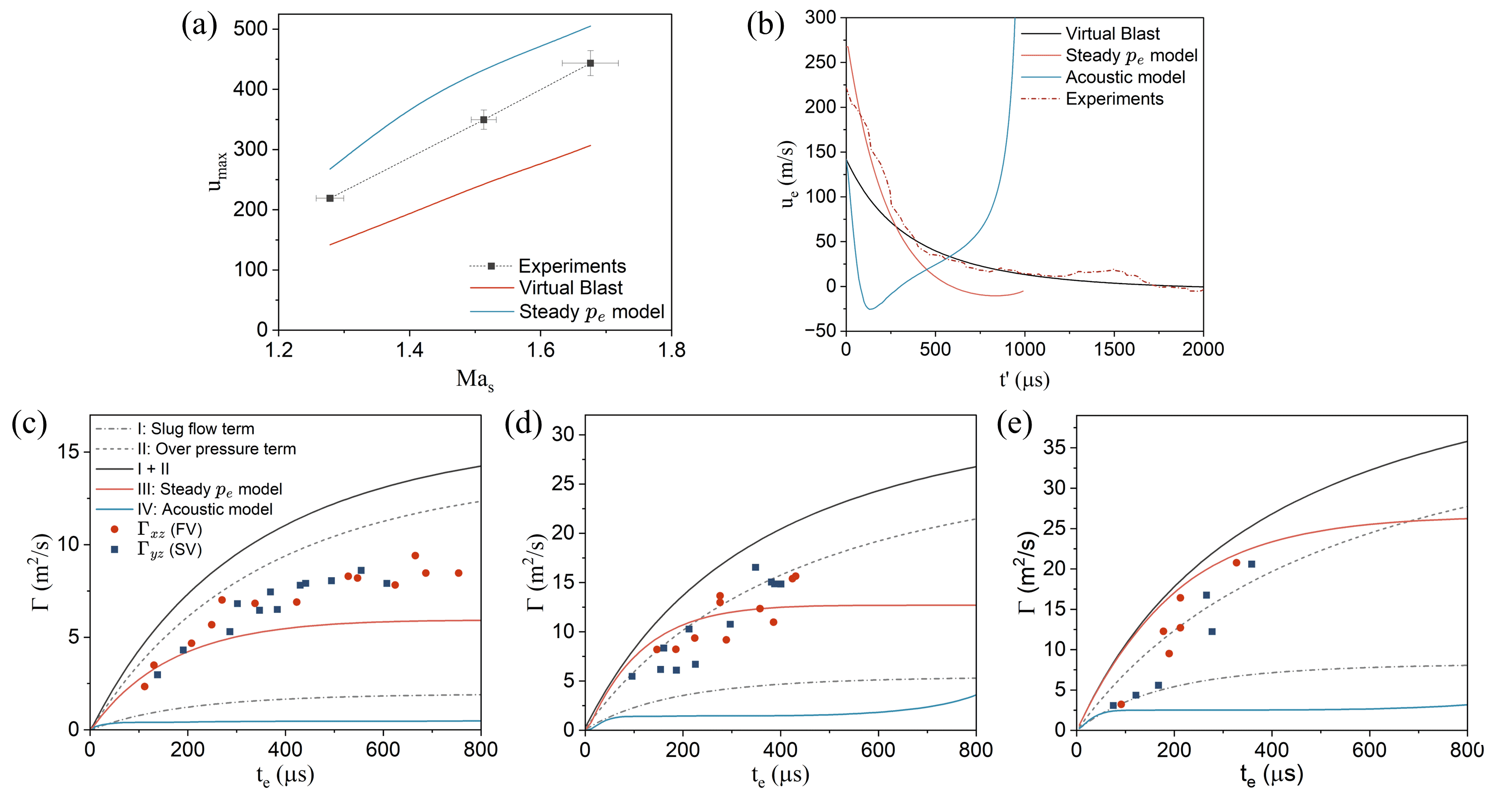}}
  \caption{(a) Comparison of peak velocities obtained from the experiments with virtual blast and steady pressure models. (b) Temporal variation of the axial flow velocity and comparison with various models, where $t'$ represents time starting from the appearance of peak velocity. Temporal evolution of circulation estimated along two principal planes determined from PIV fields and comparison with Slug flow based models considering tube exit condition: I. Virtual blast with slug flow term; II. Virtual blast with overpressure term; III. Steady pressure model IV. Acoustic model (c) $V_c = 6kV$ (d) $V_c = 8kV$ (e) $V_c = 10kV$}
\label{fig:circulation}
\end{figure}

Before moving to the vortex dynamics, we compare the predictions of the model with experiments in terms of the axial flow velocity. The peak velocity values are compared in figure~\ref{fig:circulation}a. The theoretical peak velocity closely aligns with the observed experimental values, specifically when using the closest downstream measurement. This measurement reflects a lower value compared to the source, accounting for decay and roll-up effects associated with CVR and trailing jet. This approach is adopted because the data near the tube opening, particularly during the initial stages, are less accurate due to the intense flash produced by the explosion, as discussed earlier. Furthermore, the time series data of the axial flow is compared with the models deduced in the present study in figure~\ref{fig:circulation}b, where $t'=t-t_{\rm ref}$ is used with reference time $t_{\rm ref}$ corresponding to the appearance of peak velocity. We use $z/D_e=1$ data from the experiments and find that the steady $p_e$ model predicts the observations most accurately, including other observations discussed earlier.

The vortex rings are characterized by circulation $\Gamma$, an integral quantity of vorticity $\omega = \nabla \times u $, representative of the strength of rotation. For compressible flows, the vorticity equation can be written as \citep{thompson1972compressible, green2012fluid}
\begin{eqnarray}
    \frac{D\omega}{Dt}  & = & (\omega \cdot \nabla) u -\omega (\nabla \cdot u) + \nabla T \times \nabla s  \nonumber\\
    & + & \frac{4}{3}\mu \left\{ \nabla \left(\frac{1}{\rho}\right) \times \nabla (\nabla \cdot u) \right\} + \mu \left\{\frac{1}{\rho}\nabla^2\omega - \nabla \left(\frac{1}{\rho}\right) \times (\nabla \times \omega ) \right\}  \label{eq:vorticity}
\end{eqnarray}

Momentarily, neglecting the terms associated with compressibility effects and integrating this equation for an axisymmetric system invoking the slug approximation gives \citep{kruegerOverpressureCorrectionSlug2005,xiangCirculationProductionModel2023}
\begin{equation}
    \frac{d\Gamma}{dt} = \frac{1}{2} u_{cl}^2+\frac{\Delta p_{cl}}{\rho_{cl}}
\end{equation}

which is the slug flow model with an overpressure correction term with $\Delta p_{cl}=p_{cl}-p_0$, which is an artifact of non-parallel flow at the nozzle opening \citep{kruegerOverpressureCorrectionSlug2005}. This equation can be integrated to obtain
\begin{equation} \label{eq:slug}
    \Gamma=\Gamma_u+\Gamma_p=\int_{0}^{t}{\frac{1}{2}u_{cl}^2dt}+\int_{0}^{t}\frac{\Delta p_{cl}}{\rho_{cl}}dt
\end{equation}

We will consider this simplified equation to determine the theoretical circulation production by imposing $u_{cl} = u_{e},\ p_{cl}=p_0$ (no overpressure) for the model with steady outlet pressure, and $u_{cl} = u_{e},\ p_{cl}=p_e$ for the virtual blast model and acoustic model. The implications of these drastic assumptions and contributions from the compressibility effects will be discussed subsequently. 
The slipstream model proposed by \citet{sunVorticityProductionShock2003} for vorticity generated by a diffracting shock wave, however, entails the same relation as equation~\ref{eq:slug} without the overpressure term, since $p_{cl}=p_0$.

The predictions from the models are compared with the experimental observations as illustrated in figure~\ref{fig:circulation}c-e, where the circulation is estimated experimentally over two principal planes $xz$ and $yz$, aligned with the FV and SV, respectively. The circulation measurement in both these planes gives similar values as have been observed in earlier studies involving asymmetric incompressible vortex rings \citep{ofarrellPinchoffNonaxisymmetricVortex2014,steinfurthVortexRingsProduced2020}. 
The model with a steady outlet pressure closely agrees with the experiments, slightly under-predicting the values. This is due to the other sources of circulation production arising from compressibility effects, including the embedded shocks, slip streams, and similar features giving rise to entropy gradients.

\subsubsection{A note on assumptions and boundary conditions}
This section examines the various underlying assumptions employed in the current problem formulation and delineates the corresponding limitations. Firstly, we assumed the flow to be one-dimensional for the shock motion and flow evolution at the exit. The shock motion inside the tube involves a transition from a cylindrical geometry to a planar one through multiple reflections and wave steepening. This initial interaction of the blast wave with the tube walls is a complex phenomenon that we cannot observe due to the intense light. However, there was a minimal influence of this effect on the modeling of the blast wave using a planar geometry assumption, which indicates that this initial interaction is immediate and lasts for a short period. This effect was presumably accounted for by considering a virtual source with tube opening at $r = L_{\rm eff}>L$. Also, the blast wave outside the shock tube is approximated to be a cylindrical wave; however, the exact geometry is an intermediate between the cylindrical and spherical shapes. We justified the role of a shorter span of the shock tube ($D_e=D_2$) as a dominant mechanism; however, a better strategy is necessary for blast waves with intermediate strengths with an asymmetric source of explosion \citep{chiuBlastWavesAsymmetrical1977, svetsovJetVortexFlow1997}.

Later, to evaluate the outflow and the reflected flow near the exit, we implemented the MOC and Riemann invariants for a one-dimensional flow system. This assumption is validated by earlier studies where a one-dimensional flow was observed in the proximity of the wall through the interferogram, in which the fringes were perpendicular to the wall \citep{sunFormationSecondaryShock1997}. We extended this flow to the centerline, assuming a uniform flow across the cross-section. Thus, the one-dimensional model presented here overlooks the unsteady boundary layer at the tube walls; however, this assumption is justified given the extremely short time scales associated with the flow. A prominent shear layer originates at the corner outside the shock tube during vortex formation, and the flow decay leads to destabilization of the shear layer and eventually flow separation at the walls during the flow reversal or inflow (see supplementary material). Beyond this instant, the one-dimensional flow and isentropic flow assumptions are invalid \citep{rudingerReflectionPressureWaves1957}.

The steady exit pressure model has an intrinsic flaw of assuming a steady state pressure at the exit plane for an extremely transient outflow with significant incident overpressure. It is expected to observe an overpressure in response to this incident impulse. Although this was considered in a model based on acoustics, the observation does not match well with the predictions. The acoustic model may suffer from the linearized consideration of a flow that is actually non-linear. Furthermore, applying a steady-state pressure boundary condition results in a supersonic flow at the tube exit, which presents a significant crisis for the method of characteristics. The incident subsonic flow is instantaneously set to a supersonic condition through pressure normalization. This supersonic outflow at the exit entails that the reflected characteristics will not exist inside the tube, although an upstream point just beyond the exit plane (with subsonic flow) will give rise to left-running characteristics within the tube. So the model in the present form indicates the absence of the immediate expansion wave following the incident shock for $V_c\geq 7kV$, which is clearly not the case. This "existential crisis" of the left-running characteristics during the supersonic outflow is an open question if the steady-state pressure assumption at the boundary is followed. Evidently, a different boundary condition is necessary in the early stages with overpressure. A non-linear response function to the incident characteristics that incorporates the pressure history is required to assess the flow, where eventually, the pressure at the tube exit approaches the ambient pressure asymptotically.

The angular position of the PM fan $\beta$ and shear layer $\theta$ align with the experimental trends temporally, but there is a consistent offset observed. This deviation might originate from the deployment of a quasi-steady assumption in using standard steady-state PM fans solutions, or it is evidence of the presence of overpressure at the tube opening. Furthermore, the shear layer has a finite thickness due to microscale mixing \citep{rikanatiShockWaveMachReflectionSlipStream2006}, leading to a finite thickness of the slip stream, which might be a reason for the measurement error in $\theta$. The shear layer evolution is also coupled to the early growth of the CVR, which is not considered here. We also assume the tail of the PM fan to be at the tube exit plane, referring to previous literature; however, there are no experimental observations in the present study to confirm this assumption.

Further, we used an axisymmetric slug model to deduce circulation production, which is an extreme oversimplification of the present scenario. There is a scarcity of such models for vortex ring formation from asymmetric ducts, nonetheless accounting for the compressibility effects in the same. \cite{virkCompressibleVortexReconnection1995} presented an elaborate argument on the circulation production mechanisms during a compressible vortex reconnection. This basically involved the re-interpretation and estimation of the circulation production from the rest of the term, which we neglected in equation~\ref{eq:vorticity}. They demonstrated that the formation of shocklets during this process played a significant role, which is analogous to the embedded shock formation observed in the current problem. Cumulatively, neglecting such effects leads to an underestimation of the CVR strength. 

\section{Conclusion}
In this study, the dynamics of unsteady flow at the shock tube opening established through a blast wave is investigated in detail.  A shock tube employing the wire-explosion technique with a rectangular cross-section is used to generate blast waves over a Mach number range of $Ma_{se}=1.2–1.8$.  A comprehensive report is presented on the associated flow features, including the primary blast wave and the diffracted blast wave outside. This also includes the formation of other transient flow features, which include compressible vortex ring (CVR), expansion fan, reverse shock or reshock, embedded shocks, shear layer, and Prandtl-Meyer expansion fan. An approximate solution based on energy integral and a power law density profile has been implemented for the blast system inside and outside the shock tube, where the equivalent blast source is deduced from the experimental shock trajectories. This is accompanied by the fact that only $\sim 1\%$ of the blast energy is transferred to the blast system outside the shock tube, while the rest of it is presumably consumed in the CVR formation and secondary shock features. The interaction of the blast wave with the tube opening is resolved using various approaches. The simplest one involved neglecting the tube opening altogether. The other approaches considered a boundary condition at the tube exit, and the method of characteristics (MOC) was deployed to solve the interaction between the incoming and reflected characteristics, including the expansion wave. The choices of a boundary condition involved either a steady pressure at the tube opening or an impulse response to the incident over-pressure based on acoustic considerations. The steady pressure outlet model enabled the prediction of various experimental observations more accurately, including phenomena of a supersonic efflux and subsequent formation of the embedded shock, PM expansion fan, and shear layer. It also predicted the formation of the reshock through a wave steepening mechanism, axial flow velocity of the induced gas motion and circulation production associated with the CVR. The key assumptions and their implications have been thoroughly discussed, highlighting potential areas for improvement in future studies. Various observations documented in this study were not reported earlier, as the blast wave produced an unsteady incident flow that interacted with the open-ended shock tube and resulted in the formation of previously unseen reshock and embedded shock shedding. The specific mechanism responsible for these features warrants further investigation. Additionally, there is a need to establish a more precise boundary condition at the tube outlet that is appropriate for such highly transient flow conditions.


\backsection[Acknowledgements]{S.J.R. and A.A. would like to thank the Prime Minister Research Fellowship (PMRF) for the financial support. S.B. would like to acknowledge the support from the Indian National Academy of Engineering (INAE) Chair professorship.}


\backsection[Declaration of interests]{The authors report no conflict of interest.}


\backsection[Author ORCIDs]{S. J. Rao, https://orcid.org/0000-0001-6539-5814; A. Aravind, https://orcid.org/0000-0002-5499-5408; S. Basu, https://orcid.org/0000-0002-9652-9966}


\appendix

\section{Closed form solutions in the Strong Shock limit}\label{appA}

In this section, we deduce closed-form solutions in the strong shock limit and obtain suitable scales for the problem. The resulting solutions are not exactly valid in the range of Mach numbers in the present study, however, they give us a useful insight relevant to the problem.

In the strong shock limit, we have $M_s \gg1$, we can assume $\eta=1/M_s^2 \rightarrow 0$. As a consequence, the terms on the LHS of equations \ref{eq:mass} and \ref{eq:momentum} vanish, simplifying the system in terms of only the independent parameter $\xi = r/R_s(t)$. The boundary condition at the shock front $\xi = 1$ arising from Rankine-Hugoniot conditions simplifies to

\begin{eqnarray}
    \phi\left(1\right)= f\left(1\right)=\frac{2}{\gamma+1},\ \ \ 
\psi\left(1\right)=\frac{\gamma+1}{\gamma-1}
\end{eqnarray}

A self-similar solution to this set of equations exists \citep{lee2016gas} which depends only on the similarity parameter $\xi$, where the decay parameter $\theta$ is now a constant. Based on the equation \ref{eq:theta}, we have
\begin{equation}
    {\dot{R}}_s = CR_s^\theta
\end{equation}
 where $C$ is a constant of integration and we can say that the blast trajectory follows
\begin{equation}
    R_s = At^N
\end{equation}
 where $A$ and $N$ are constant. The obtained self-similar solution, and dependency of the coefficients on the geometry parameter $j$ is as below \citep{lee2016gas}

\begin{equation}
    \phi(\xi) = \phi(1)\xi
\end{equation}
\begin{equation}
    \psi(\xi) = \psi(1)\xi^q
\end{equation}
\begin{equation}
    f(\xi) = f(1)+\frac{\psi(1)\phi(1)}{q+2} \left( 1-\phi(1)-\theta \right) \left(\xi^{q+1}-1\right)
\end{equation}

where

\begin{equation}
    \theta = -\frac{j+1}{2},\ \ N=\frac{1}{1-\theta},\ \ A = \{C(1-\theta)\}^N,\ \ q = (j+1)\left( \psi(1)-1 \right)
\end{equation}

In the dimensional form these can be rewritten as 

\begin{equation}
    u=\hat{u}\frac{r}{t}
\end{equation}

\begin{equation}
    \psi = \hat{\rho} r^q t^{-Nq}
\end{equation}

\begin{equation}
    p = \hat{p}_1 t^{2(N-1)}+\hat{p}_2 r^{q+2} t^{-Nq-2}
\end{equation}

\begin{equation}
    \hat{u} = N\phi(1),\quad \hat{\rho}=\frac{\rho_0\psi(1)}{A^q},\quad \hat{p}_1=f(1)\rho_0 A^2 N^2 (1-F), \quad \hat{p}_2= \frac{f(1)\rho_0 N^2 F}{A^{q}}
\end{equation}

where
\begin{equation}
    F = \frac{\psi(1)\phi(1)}{f(1)} \frac{(1-\phi(1)-\theta)}{q+2}
\end{equation}

The normalized energy integral consists of the term $I$ \citep{bachAnalyticalSolutionBlast1970} given by 

\begin{eqnarray}
    I &=& \int_{0}^1 \left(\frac{f}{\gamma-1} +\frac{\psi\phi^2}{2} \right)\xi^jd\xi \nonumber \\
    &=& \frac{f(1)(1-F)}{(\gamma-1)(j+1)} +\frac{1}{q+j+3} \left\{ \frac{f(1)F}{\gamma-1} +\frac{\psi(1)\phi^2(1)}{2} \right\}
\end{eqnarray}

which can then be used to estimate the coefficient associated with the shock trajectory as

\begin{equation}
    C = \sqrt{\frac{E_0}{\rho_0}\frac{1}{k_jI}} = \sqrt{\frac{c_0^2 R_0^{j+1}}{I}}
\end{equation}

The flow associated with these high Mach number shocks will generate a supersonic flow and will generate a Prandtl-Meyer expansion fan outside the shock tube instead of the inward-moving expansion wave that we observe in our experiments. However, we'll use this solution to determine the circulation production from the slug term, with the objective of determining suitable scales and presenting a representative calculation. Readers can integrate the over-pressure term as well and also incorporate the PM fan through isentropic relations alongside the relevant pressure outlet boundary condition. At the tube exit, we consider the flow based on the virtual blast model discussed earlier (i.e., without the effect of opening). For the tube opening $r=L$, so substituting the same, we get

\begin{equation}
    u(r=L,t) = \hat{u} \frac{L}{t}
\end{equation}

And circulation production, assuming the tube centerline velocity $u_{cl}(t) = u(r=L,t)$ is given by 

\begin{equation}
    \frac{d\Gamma}{dt} = \frac{1}{2}u_{cl}^2 = \frac{1}{2}\hat{u}^2\frac{L^2}{t^2}
\end{equation}

integrating, beginning from the shock arrival time $t_{se}$, we get

\begin{equation}
    \Gamma = \int_{t_{se}}^t \frac{1}{2}\hat{u}^2\frac{L^2}{t'^2} dt' \nonumber
\end{equation}
\begin{equation}
    \Gamma = \frac{1}{2}\hat{u}^2 L^2 \left( \frac{1}{t_{se}}-\frac{1}{t} \right)
\end{equation}

as $t\rightarrow\infty, \quad \Gamma\rightarrow \Gamma_{\rm max}$, hence

\begin{equation}
    \Gamma_{\rm max} = \frac{1}{2} \frac{\hat{u}^2 L^2}{t_{se}}
\end{equation}

defining $\bar{\tau}= t/t_{se}$, we have
\begin{equation}
    \frac{\Gamma}{\Gamma_{\rm max}} = 1-\frac{1}{\bar{\tau}}
\end{equation}

Now, considering the definition of normalized time, often reported as formation time in the vortex dynamics literature \citep{rosenfeldCirculationFormationNumber1998}, we have

\begin{equation}
    t^\ast = \frac{L_u(t)}{D_e} = \frac{\int_{t_{se}}^t u_{cl} dt}{D_e}
\end{equation}

where $L_u$ is the equivalent piston displacement in terms of the volume flux of fluid and $D_e$ is the tube diameter, which in this case will be taken as $D_e = D_2$. Substituting and simplifying, we get
\begin{equation}
    t^\ast = \frac{L}{D} \hat{u} \ln{\bar{\tau}}
\end{equation}

The aforementioned analysis characterizes \( t_{se} \) as a pertinent temporal scale associated with the outflow dynamics. Depending on the specific scenario, particularly during extreme explosive events, this approach facilitates the derivation of closed-form analytical solutions.

\bibliographystyle{jfm}
\bibliography{jfm}








\end{document}